\documentclass[review]{elsarticle}
\usepackage{lineno,hyperref}
\usepackage[english]{babel}
\usepackage{color}
\usepackage{amsmath}
\usepackage{amsfonts}
\usepackage{dsfont}
\usepackage{amssymb}
\usepackage{makeidx}
\usepackage{graphicx}
\usepackage{float} 
\usepackage{lmodern}
\usepackage{hyperref}
\usepackage{amsthm}
\usepackage{mathabx}
\usepackage{fancyhdr}
\usepackage{caption}
\usepackage{todonotes}
\usepackage{subcaption}
\usepackage[normalem]{ulem}
\usepackage{soul,ulem}
\usepackage{natbib}
\usepackage{yfonts}
\usepackage{makecell}
\usepackage{booktabs,dcolumn}
\usepackage[toc,page]{appendix}

\usepackage[a4paper, top=3cm, bottom=3cm, left=2.5cm, right=2.5cm]{geometry}

\theoremstyle{definition}

\newtheorem{rem}{Remark}
\newtheorem{exmp}{Example}[section]

\newtheorem{definition}{Definition}[section]

\begin{document}

\begin{frontmatter}

\title{Spatiotemporal large-scale networks shaped by air mass movements}

\author[add1]{Maria Choufany}
\ead{maria.choufany@inrae.fr}
\author[add1]{Davide Martinetti}
\ead{davide.martinetti@inrae.fr}
\author[add1]{Rachid Senoussi}
\ead{rachid.senoussi@inrae.fr}
\author[add2]{Cindy E. Morris}
\ead{cindy.morris@inrae.fr}
\author[add1]{Samuel Soubeyrand}
\ead{samuel.soubeyrand@inrae.fr}

\address[add1]{INRAE, BioSP, 84914 Avignon, France}
\address[add2]{INRAE, UR407 Pathologie V{\'e}g{\'e}tale, Centre de Recherche PACA, Montfavet, France}

\begin{abstract}

The movement of atmospheric air masses can be seen as a continuous and complex flow of particles hovering over our planet. It can however be locally simplified by considering three-dimensional trajectories of air masses connecting distant areas of the globe during a given period of time.

In this paper, we present a mathematical framework to construct spatial and spatiotemporal networks where the nodes are the subsets of a partition of a geographical area and the links between these nodes are inferred from sampled trajectories of air masses passing over and across the nodes.
We propose different estimators of link intensities relying on different bio-physical hypotheses and covering adjustable time periods. This approach leads to a new class of spatiotemporal networks characterized by adjacency matrices. We applied the approach in two real geographical contexts: the watersheds of the French region Provence-Alpes-Côte d'Azur and the coastline of the Mediterranean Sea. The analysis of the constructed networks allowed identifying a marked seasonal pattern in air mass movements in the study areas.

These constructed networks can be used to investigate issues, e.g., in aerobiology and epidemiology of airborne plant pathogens. Similar networks could be estimated from other types of trajectories, such as animal trajectories.

\end{abstract}

\end{frontmatter}
\thispagestyle{empty}

\section{Introduction}

Atmospheric air masses are volumes of air with a defined temperature and water vapor content that have long been known to rule fundamental atmospheric phenomena like weather and air currents. Their composition is mostly inert gases, but both organic and inorganic particles have been found to linger in high-altitude air as a consequence of the constant interaction of air masses with the earth's surface below them. A non-exhaustive list includes gases and minerals like wildfire smoke, radioactive material, dust, sand, volcanic ash and sea salt, but also living organisms such as pollen, fungal spores, bacteria, virus and small insects. Despite the relative sparse density of these particles with respect to the volume of an air mass, their presence and transportation across the planet has proven to have strong effects on many phenomena impacting human health and safety (pollen \citep{mahura2007elevated, vsauliene2006application, bogawski2019detecting}, dust concentrations \citep{khaniabadi2017impact, aciego2017dust}, nuclear byproducts \citep{moroz2010predictions, rolph2014modeling}, human, animal and plant epidemics \citep{leyronas2018assessing,wang2010long,aylor1990role,mundt2009aerial,sadys2014back,hiraoka2017seasonal}, air pollution \citep{liu2018seasonal, liu2018study, talbi2018assessment}, and rainfall \citep{chen2018analysis, armon2018synoptic, rabinowitz2019investigation}). 

The rise in the number of publications on these subjects suggests a growing interest of the scientific community on the effects of air-mass movements on the biosphere, that has surely been boosted by recent available developments, such as the Hybrid Single-Particle Lagrangian Integrated Trajectory model (\verb"HYSPLIT", \cite{stein2015noaa}), allowing reconstruction of actual air-mass movements at rather fine geographical and temporal scales and with a global cover. 

The vast majority of studies focused on isolated events, such as dust storms or peaks of air pollutants, that are rather concentrated in time (from few hours to few weeks) and/or space (just a few locations such as cities). Nonetheless, the movement of air masses is expected to have impacts on a broader spatiotemporal scale, as reviewed in recent studies \citep{leyronas2018assessing,margosian2009connectivity}. The purpose of the present paper is then to propose a mathematical framework for studying air-mass movements on large spatiotemporal scales, under the hypothesis that these movements can create stable and recurrent connections between distant portions of a territory. The very nature of these connections will be further specified throughout the manuscript, but as a general rule we will consider that any pair of points (or areas) in space can have a certain degree of connection, regardless of their geographic distance, provided that there are recurrent air-mass trajectories that connect the two points (or areas). The direction and strength of these connections will be estimated by looking at the trajectories linking every pair of points/areas and weighting them according to appropriate measures. In this perspective, it seems natural to resort to graph and network theory, since the formalism of nodes and edges provides an adequate environment for describing complex connections and can further be used to deepen into the topology of the constructed networks in order to infer interesting properties of the graphs, such as the presence of hubs.

From a generic statistical point of view, we aim to (i) estimate the weighted and directed edges of a graph using a sample of trajectories of {\it individuals} traveling through the space formed by the nodes of the graph, and (ii) characterize the estimated graph based on relevant statistics.

In the following sections, we first introduce the definitions and properties that will allow us to describe and then estimate connections between points/areas in space via spatiotemporal trajectories. Then, we propose several types of measures to model diverse types of connections. The expected output consists of a spatiotemporal graph describing the network of links induced by trajectories. It's worth noting that our approach is meant to infer connectivity induced by air-mass movements and it is readily applicable to \verb"HYSPLIT"-type data, but we have maintained a sufficient level of generality to be applied to other phenomena, provided that trajectory data are available  (e.g. animal trajectories). Finally, we apply our method to two case studies concerning the coastline of the Mediterranean sea and the French region of Provence-Alpes-C\^{o}te d'Azur. The two case studies have different spatiotemporal granularities and they will be used to provide examples of application of the proposed methodology.

\section{Framework for the definition of trajectory-based networks}\label{subsec:traj_to_net}
\label{theo_approach}

In this section we show how a set of trajectories evolving within space during a finite time interval can be used to construct pertinent spatiotemporal networks. We first recall some basic definitions related to networks (Section~\ref{subsec:net_theory}) and then propose a statistical methodology to infer the network structure from a data set of trajectories (Section~\ref{subsec:traj_to_net}).

\subsection{Network theory}
\label{subsec:net_theory}

Network theory (a.k.a. graph theory) is a mathematical formalism introduced by Leonhard Euler to describe the famous K{\"o}nigsberg bridge problem \citep{newman2003structure,strogatz2001exploring,west1996introduction}. The two basic components of a network are a set of \textit{nodes} linked by a set of edges. Nodes can represent a variety of things, such as persons, regions, computers, neurons, etc., while edges are used to describe the connections between those nodes. Formally, a \textit{network} $G=\left( V, E\right)$ is defined as a set of \textit{nodes} (or vertices) $V=\lbrace v_1,v_2,\ldots ,v_N \rbrace$ connected by a set of \textit{edges} $E= \lbrace e_{ij} \rbrace_{i,j \in \lbrace 1,\ldots,N \rbrace}$. 
A natural way of representing a network is given by means of a $N\times N$ square matrix $M$, usually referred to as an \textit{adjacency matrix}, whose term $(i,j)$, $M_{ij}$, is non-zero an edge exists between $i$ and $j$. 
By convention, adjacency matrices are defined to have an empty diagonal (i.e. $M_{ii}=0, i\in\lbrace 1,\ldots ,N\rbrace$), meaning that nodes cannot be self connected. If $M$ is symmetrical (i.e. $M_{ij}=M_{ji}, i,j \in\lbrace 1,\ldots ,N\rbrace$), then the network is said to be \textit{undirected}, and \textit{directed} otherwise. If $M_{ij}\in \lbrace{0,1}\rbrace$, the network is said to be \textit{binary}, meaning that an edge between two nodes $i$ and $j$ either exists or does not. Otherwise, if $M_{ij}\in \mathbb{R}$, the network is said to be \textit{weighted}, meaning that the edge between nodes $i$ and $j$ are more or less connected.

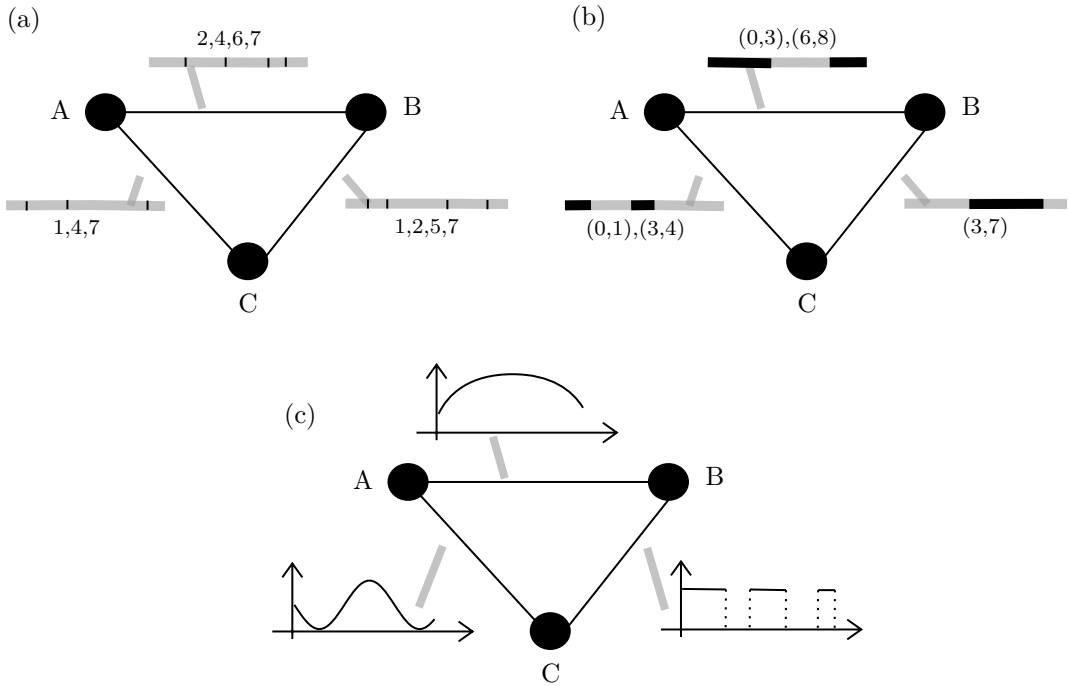
\begin{figure}[H]
    \centering
\tikzset{every picture/.style={line width=0.75pt}} 
\begin{tikzpicture}[x=0.75pt,y=0.75pt,yscale=-1,xscale=1]
\draw  [fill={rgb, 255:red, 0; green, 0; blue, 0 }  ,fill opacity=1 ] (80,96) .. controls (80,91.03) and (84.37,87) .. (89.75,87) .. controls (95.13,87) and (99.5,91.03) .. (99.5,96) .. controls (99.5,100.97) and (95.13,105) .. (89.75,105) .. controls (84.37,105) and (80,100.97) .. (80,96) -- cycle ;
\draw  [fill={rgb, 255:red, 0; green, 0; blue, 0 }  ,fill opacity=1 ] (210,96) .. controls (210,91.03) and (214.37,87) .. (219.75,87) .. controls (225.13,87) and (229.5,91.03) .. (229.5,96) .. controls (229.5,100.97) and (225.13,105) .. (219.75,105) .. controls (214.37,105) and (210,100.97) .. (210,96) -- cycle ;
\draw  [fill={rgb, 255:red, 0; green, 0; blue, 0 }  ,fill opacity=1 ] (151,171) .. controls (151,166.03) and (155.37,162) .. (160.75,162) .. controls (166.13,162) and (170.5,166.03) .. (170.5,171) .. controls (170.5,175.97) and (166.13,180) .. (160.75,180) .. controls (155.37,180) and (151,175.97) .. (151,171) -- cycle ;
\draw    (89.75,96) -- (153.5,166) ;
\draw    (89.75,96) -- (210,96) ;
\draw    (165.5,174) -- (219.75,105) ;
\draw [color={rgb, 255:red, 155; green, 155; blue, 155 }  ,draw opacity=0.6 ][line width=3.75]    (40.07,142.87) -- (84.78,142.73) -- (119.42,143.27) ;
\draw [color={rgb, 255:red, 155; green, 155; blue, 155 }  ,draw opacity=0.6 ][line width=3.75]    (209.5,142) -- (261.1,142) -- (290.73,142.2) ;
\draw [color={rgb, 255:red, 155; green, 155; blue, 155 }  ,draw opacity=0.6 ][line width=3.75]    (111.4,70.87) -- (167.1,71) -- (190.73,70.87) ;
\draw [color={rgb, 255:red, 155; green, 155; blue, 155 }  ,draw opacity=0.6 ][line width=3]    (132.1,73.2) -- (138.1,94.2) ;
\draw [color={rgb, 255:red, 155; green, 155; blue, 155 }  ,draw opacity=0.6 ][line width=3]    (102.1,143) -- (107.1,128.2) ;
\draw [color={rgb, 255:red, 155; green, 155; blue, 155 }  ,draw opacity=0.6 ][line width=3]    (221.1,142.2) -- (209.1,128.2) ;
\draw    (129.73,68.47) -- (129.73,73.13) ;
\draw    (171.07,68.8) -- (171.07,73.47) ;
\draw    (149.73,68.47) -- (149.73,73.13) ;
\draw    (179.73,68.47) -- (179.73,73.13) ;
\draw    (110.73,140.47) -- (110.73,145.13) ;
\draw    (70.73,140.13) -- (70.73,144.8) ;
\draw    (50.4,140.8) -- (50.4,145.47) ;
\draw    (220.73,139.8) -- (220.73,144.47) ;
\draw    (230.4,139.8) -- (230.4,144.47) ;
\draw    (260.4,140.13) -- (260.4,144.8) ;
\draw    (280.4,140.13) -- (280.4,144.8) ;
\draw  [fill={rgb, 255:red, 0; green, 0; blue, 0 }  ,fill opacity=1 ] (359,96) .. controls (359,91.03) and (363.37,87) .. (368.75,87) .. controls (374.13,87) and (378.5,91.03) .. (378.5,96) .. controls (378.5,100.97) and (374.13,105) .. (368.75,105) .. controls (363.37,105) and (359,100.97) .. (359,96) -- cycle ;
\draw  [fill={rgb, 255:red, 0; green, 0; blue, 0 }  ,fill opacity=1 ] (489,96) .. controls (489,91.03) and (493.37,87) .. (498.75,87) .. controls (504.13,87) and (508.5,91.03) .. (508.5,96) .. controls (508.5,100.97) and (504.13,105) .. (498.75,105) .. controls (493.37,105) and (489,100.97) .. (489,96) -- cycle ;
\draw  [fill={rgb, 255:red, 0; green, 0; blue, 0 }  ,fill opacity=1 ] (430,171) .. controls (430,166.03) and (434.37,162) .. (439.75,162) .. controls (445.13,162) and (449.5,166.03) .. (449.5,171) .. controls (449.5,175.97) and (445.13,180) .. (439.75,180) .. controls (434.37,180) and (430,175.97) .. (430,171) -- cycle ;
\draw    (368.75,96) -- (432.5,166) ;
\draw    (368.75,96) -- (489,96) ;
\draw    (444.5,174) -- (498.75,105) ;
\draw [color={rgb, 255:red, 155; green, 155; blue, 155 }  ,draw opacity=0.6 ][line width=3.75]    (319.07,142.87) -- (363.78,142.73) -- (398.42,143.27) ;
\draw [color={rgb, 255:red, 155; green, 155; blue, 155 }  ,draw opacity=0.6 ][line width=3.75]    (488.5,142) -- (540.1,142) -- (569.73,142.2) ;
\draw [color={rgb, 255:red, 155; green, 155; blue, 155 }  ,draw opacity=0.6 ][line width=3.75]    (390.4,70.87) -- (446.1,71) -- (469.73,70.87) ;
\draw [color={rgb, 255:red, 155; green, 155; blue, 155 }  ,draw opacity=0.6 ][line width=3]    (411.1,73.2) -- (417.1,94.2) ;
\draw [color={rgb, 255:red, 155; green, 155; blue, 155 }  ,draw opacity=0.6 ][line width=3]    (381.1,143) -- (386.1,128.2) ;
\draw [color={rgb, 255:red, 155; green, 155; blue, 155 }  ,draw opacity=0.6 ][line width=3]    (500.1,142.2) -- (488.1,128.2) ;
\draw [color={rgb, 255:red, 0; green, 0; blue, 0 }  ,draw opacity=1 ][line width=3.75]    (390.4,70.87) -- (422.07,71.4) ;
\draw [color={rgb, 255:red, 0; green, 0; blue, 0 }  ,draw opacity=1 ][line width=3.75]    (451.23,70.93) -- (469.73,70.87) ;
\draw [color={rgb, 255:red, 0; green, 0; blue, 0 }  ,draw opacity=1 ][line width=3.75]    (319.07,142.87) -- (332.07,142.73) ;
\draw [color={rgb, 255:red, 0; green, 0; blue, 0 }  ,draw opacity=1 ][line width=3.75]    (352.12,142.87) -- (363.78,142.73) ;
\draw [color={rgb, 255:red, 0; green, 0; blue, 0 }  ,draw opacity=1 ][line width=3.75]    (520.92,142.23) -- (557.92,142.1) ;
\draw  [fill={rgb, 255:red, 0; green, 0; blue, 0 }  ,fill opacity=1 ] (231,282) .. controls (231,277.03) and (235.37,273) .. (240.75,273) .. controls (246.13,273) and (250.5,277.03) .. (250.5,282) .. controls (250.5,286.97) and (246.13,291) .. (240.75,291) .. controls (235.37,291) and (231,286.97) .. (231,282) -- cycle ;
\draw  [fill={rgb, 255:red, 0; green, 0; blue, 0 }  ,fill opacity=1 ] (361,282) .. controls (361,277.03) and (365.37,273) .. (370.75,273) .. controls (376.13,273) and (380.5,277.03) .. (380.5,282) .. controls (380.5,286.97) and (376.13,291) .. (370.75,291) .. controls (365.37,291) and (361,286.97) .. (361,282) -- cycle ;
\draw  [fill={rgb, 255:red, 0; green, 0; blue, 0 }  ,fill opacity=1 ] (302,357) .. controls (302,352.03) and (306.37,348) .. (311.75,348) .. controls (317.13,348) and (321.5,352.03) .. (321.5,357) .. controls (321.5,361.97) and (317.13,366) .. (311.75,366) .. controls (306.37,366) and (302,361.97) .. (302,357) -- cycle ;
\draw    (240.75,282) -- (304.5,352) ;
\draw    (240.75,282) -- (361,282) ;
\draw    (316.5,360) -- (370.75,291) ;
\draw [color={rgb, 255:red, 155; green, 155; blue, 155 }  ,draw opacity=0.6 ][line width=3]    (283.1,259.2) -- (289.1,280.2) ;
\draw [color={rgb, 255:red, 155; green, 155; blue, 155 }  ,draw opacity=0.6 ][line width=3]    (246.1,345) -- (258.1,314.2) ;
\draw [color={rgb, 255:red, 155; green, 155; blue, 155 }  ,draw opacity=0.6 ][line width=3]    (369.1,346.2) -- (360.1,315.2) ;
\draw  (173,357.2) -- (273,357.2)(183,323) -- (183,361) (266,352.2) -- (273,357.2) -- (266,362.2) (178,330) -- (183,323) -- (188,330)  ;
\draw  [color={rgb, 255:red, 0; green, 0; blue, 0 }  ,draw opacity=1 ] (184,343.8) .. controls (188.08,350.05) and (191.98,356) .. (196.5,356) .. controls (201.02,356) and (204.92,350.05) .. (209,343.8) .. controls (213.08,337.55) and (216.98,331.6) .. (221.5,331.6) .. controls (226.02,331.6) and (229.92,337.55) .. (234,343.8) .. controls (238.08,350.05) and (241.98,356) .. (246.5,356) .. controls (249.16,356) and (251.61,353.94) .. (254,350.97) ;
\draw  (245,257.2) -- (345,257.2)(255,223) -- (255,261) (338,252.2) -- (345,257.2) -- (338,262.2) (250,230) -- (255,223) -- (260,230)  ;
\draw  (367,356.2) -- (467,356.2)(377,322) -- (377,360) (460,351.2) -- (467,356.2) -- (460,361.2) (372,329) -- (377,322) -- (382,329)  ;
\draw    (377.6,335.8) -- (399.44,336.08) ;
\draw    (411.44,336.08) -- (429.44,336.4) ;
\draw    (445.44,336.4) -- (453.84,336.4) ;
\draw [line width=0.75]  [dash pattern={on 0.84pt off 2.51pt}]  (399.44,336.08) -- (399.29,355.92) ;
\draw [line width=0.75]  [dash pattern={on 0.84pt off 2.51pt}]  (411.44,336.08) -- (411.29,355.92) ;
\draw [line width=0.75]  [dash pattern={on 0.84pt off 2.51pt}]  (429.44,336.4) -- (429.29,356.24) ;
\draw [line width=0.75]  [dash pattern={on 0.84pt off 2.51pt}]  (445.44,336.4) -- (445.29,356.24) ;
\draw [line width=0.75]  [dash pattern={on 0.84pt off 2.51pt}]  (453.84,336.4) -- (453.69,356.24) ;
\draw   (256,247.8) .. controls (260.63,237.8) and (271.38,227.8) .. (292.88,227.8) .. controls (312.14,227.8) and (322.77,235.83) .. (328.1,244.69) ;
\draw (67.33,95) node  [align=left] {A};
\draw (243,93) node  [align=left] {B};
\draw (161,192) node  [align=left] {C};
\draw (151.33,60) node  [align=left] {{\footnotesize 2,4,6,7}};
\draw (75.33,155) node  [align=left] {{\footnotesize 1,4,7}};
\draw (250,154.33) node  [align=left] {{\footnotesize 1,2,5,7}};
\draw (346.33,95) node  [align=left] {A};
\draw (522,93) node  [align=left] {B};
\draw (440,192) node  [align=left] {C};
\draw (430.33,59) node  [align=left] {{\footnotesize (0,3),(6,8)}};
\draw (354.33,155) node  [align=left] {{\footnotesize (0,1),(3,4)}};
\draw (529,154.33) node  [align=left] {{\footnotesize (3,7)}};
\draw (218.33,281) node  [align=left] {A};
\draw (394,279) node  [align=left] {B};
\draw (312,378) node  [align=left] {C};
\draw (49,49.8) node  [align=left] {(a)};
\draw (331,48.8) node  [align=left] {(b)};
\draw (187,248.8) node  [align=left] {(c)};
\end{tikzpicture}
\caption{Types of temporal networks. The time of activation is indicated within the grey bar next to the edges (ranging between 0 and 8). For contact networks (a), edges activate only for one instant at the time and are marked with black vertical lines inside the grey bars. For example, in panel (a), the edge between nodes A and B is only active at instants 2, 4, 6 and 7. For interval networks (b), edges can be activated during an interval of time. For example, the edge between A and B in panel (b) is active during the time intervals (0,3) and (6,8). For contact networks (c), the edges are quantitatively more or less active across time, and the quantity of activity of any edge is described by a temporal function.}
\label{type_network}

\end{figure}

In this paper, a network is said to be \textit{spatial} \citep{barthelemy2014spatial} when nodes correspond to geographic locations, while we use the term \textit{temporal} \citep{holme2012temporal} to refer to networks where edge values can change over time. Finally, we will use the term \textit{spatiotemporal network} to refer to network that are simultaneously spatial and temporal, under the constraint that nodes cannot change position, neither appear nor disappear over time. The networks considered in this paper also fall into the rather generic definition of spatiotemporal networks. If the spatial qualifier means that the nodes of the networks represent fixed geographical locations, the temporal qualifier is more complex. Indeed, temporal networks are generally divided in the literature into two main classes, namely contact graphs or interval graphs~\citep{holme2012temporal}. The former type refers to networks where edges represents instantaneous contacts between nodes (Figure \ref{type_network}(a)), while in the second type edges are active over time intervals instead of instants of time (Figure \ref{type_network}(b)). In this paper we propose a new definition of spatiotemporal networks where nodes correspond to disjoint regions of the space and edges are computed as a function of the flow of trajectories linking these nodes (Figure \ref{type_network}(c)), as it will be explained in the rest of the current section.

\subsection{Flows and trajectory segments}\label{sec:fts}

We consider a function $\Phi: \mathbb{R} \times \mathbb{R} \times \Omega   \to  \Omega$, usually called flow on the spatial domain $\Omega$ of $\mathbb{R}^d$, satisfying the following properties:
\begin{equation}
\label{eq1}
 \left\{
\begin{array}{lcl}
\Phi(t,s,x) &=& \Phi(t,t', \Phi(t',s,x))\\
\Phi^{-1}(t,s,.) &=& \Phi (s,t,.),
\end{array}
\right. 
\end{equation}
where $s,t,t'\in\mathbb R$ and $x\in\Omega$.
For fixed $t$ and $s$, the flow $\Phi(t,s,\cdot)$ is a spatial transformation. For fixed $x$ and varying $s$ or $t$, the function gives a forward or backward trajectory of a particle over $\Omega$ between times $t \wedge s= \inf(t,s)$ and  $t \vee s = \sup(t,s)$.
If $s \leq t$, $y=\Phi(t,s,x)$ gives the future location at time $t$ of the particle presently located at $x$ at time $s$. Contrarily, if $s \geq t$, $y=\Phi(t,s,x)$ gives the location at past time $t$ that was occupied by the particle located at $x$ at present time $s$.
$\Phi(t,s,.)$ is assumed to be a bijective mapping meaning that particles following distinct trajectories cannot be at the same location at the same time.

In general, a flow is defined with respect to a possibly time-dependent vector field $F$ over $\mathbb{R} \times \Omega$, as the solution $u:\mathbb R \to \Omega$ of an ordinary differential equation (see e.g. Hamilton's equations in classical mechanics) with specified initial condition at a specified time $s$:
\begin{equation}
\centering
\left\{
\begin{array}{lcl}
\dfrac{d u}{dt}(t) &=& F (t,u(t))\\
u(s) &=& x,
\end{array}
\right.
\label{eqdiff}
\end{equation}
where $F$ is continuous and Lipschitzian over $\mathbb R \times \Omega$. In the setting introduced above, $\Phi(t,s,x)=u(t)$ with $\Phi(s,s,x)$=u(s)=$x$.
The solution $u$ represents the trajectory of the particle located at $x$ at time $s$. Varying the initial condition in System \eqref{eqdiff}, i.e. varying $s$ and $x$, leads to consider pieces of trajectories of all particles which dynamics are governed by the vector field $F$. In this article, the vector field $F$ will not be made explicit, but we will consider samples of trajectory segments (defined below) for constructing trajectory-based networks.


\begin{definition} 
\label{def:traj}
The {\it trajectory segment} associated to the flow $\Phi$ over the time interval $\Delta_{ts} = \left[ t \wedge s, t \vee s \right]$, $s,t \in \mathbb{R}$, for a particle located at $x\in\Omega$ at time $s$ is defined as follows:
\begin{equation}
\centering
\begin{array}{lcl}
\Gamma(t,s,x)&= \lbrace(t', \Phi(t',s,x)): t' \in \Delta_{ts} \rbrace. \\
\end{array}
\end{equation}
\end{definition}

If $s<t$ (resp. $s>t$), $\Gamma(t,s,x)$ is a forward (resp. backward) trajectory segment.
In this article, we are mainly interested in backward trajectories, but the framework presented here encompasses forward trajectories as well.

\begin{exmp}
\label{exemp5}
The notions of flow and trajectory segment can be adapted to cope with air mass trajectories over the Earth surface.
In this case, the spatial domain $\Omega$ representing the Earth surface is the sphere $\mathbb S^2$ in $\mathbb R^3$. If in addition, air masses are characterized by altitude and temperature evolving in space and time, then $\Omega= \mathbb S^2 \times \mathbb{R}_+ \times \mathbb{R}$, where $\mathbb{R_+}$ (resp. $\mathbb{R}$) is the domain of the altitude (resp. temperature) coordinate.

\end{exmp}

\begin{exmp}
 Animal movements and behaviour activities can also be represented with the notions of flows and trajectory segments, providing, for instance, the animal locations and the covariate value indicating whether  animals are feeding or not. In this case, $\Omega=  \mathbb{R}^2 \times \lbrace0,1 \rbrace$, where 1 stands for `the animal is feeding' and 0 otherwise. The use of a binary variable for describing the feeding activity may require the use of stochastic processes or generalized functions undergoing dynamic analog to the System \eqref{eqdiff} for constructing the flow if it is defined with respect to a vector field $F$. 
\end{exmp}

\subsection{Pointwise and integrated connectivities}\label{sec:pic}

Trajectory-based networks are grounded on the notion of {\it connectivity} used as a quantitative, directed measurement of edges between graph nodes.
In this aim, we first define the {\it pointwise connectivity} as a measure (or submeasure), in the mathematical sense, of the connectivity between a subset $A$ and a point $x$ of $\Omega$ induced by the trajectory segments $\Gamma(t,s,x)$ of a particle located at $x$ at time $s$. Then, we use the pointwise connectivity to define the {\it integrated connectivity} between two subsets $A$ and $B$ of $\Omega$ over a temporal domain $\Delta$ of $\mathbb R$ ($\Delta$ can be the union of disjoint intervals).

\begin{definition}
\label{pointwise_connectivity}
Let $x \in \Omega$ and $A \in \mathcal{B}(\Omega)$, where $ \mathcal{B}(\Omega)$ is a $\sigma$-algebra of subsets of $\Omega$. The pointwise connectivity associated to the flow $\Phi$ is defined as a real valued function $\Psi$ on $ \mathcal{B}(\Omega) \times \mathbb{R} \times \mathbb{R} \times \Omega$, conveniently denoted by $\Psi(A \mid t,s,x)$, where $A\mapsto \Psi(A \mid t,s,x)$ is a measure or a submeasure on $\Omega$ for each $t,s,x$
\end{definition}

\begin{figure}[H] 
\centering
  \begin{subfigure}[b]{0.5\linewidth}
    \centering
    \includegraphics[scale=0.5]{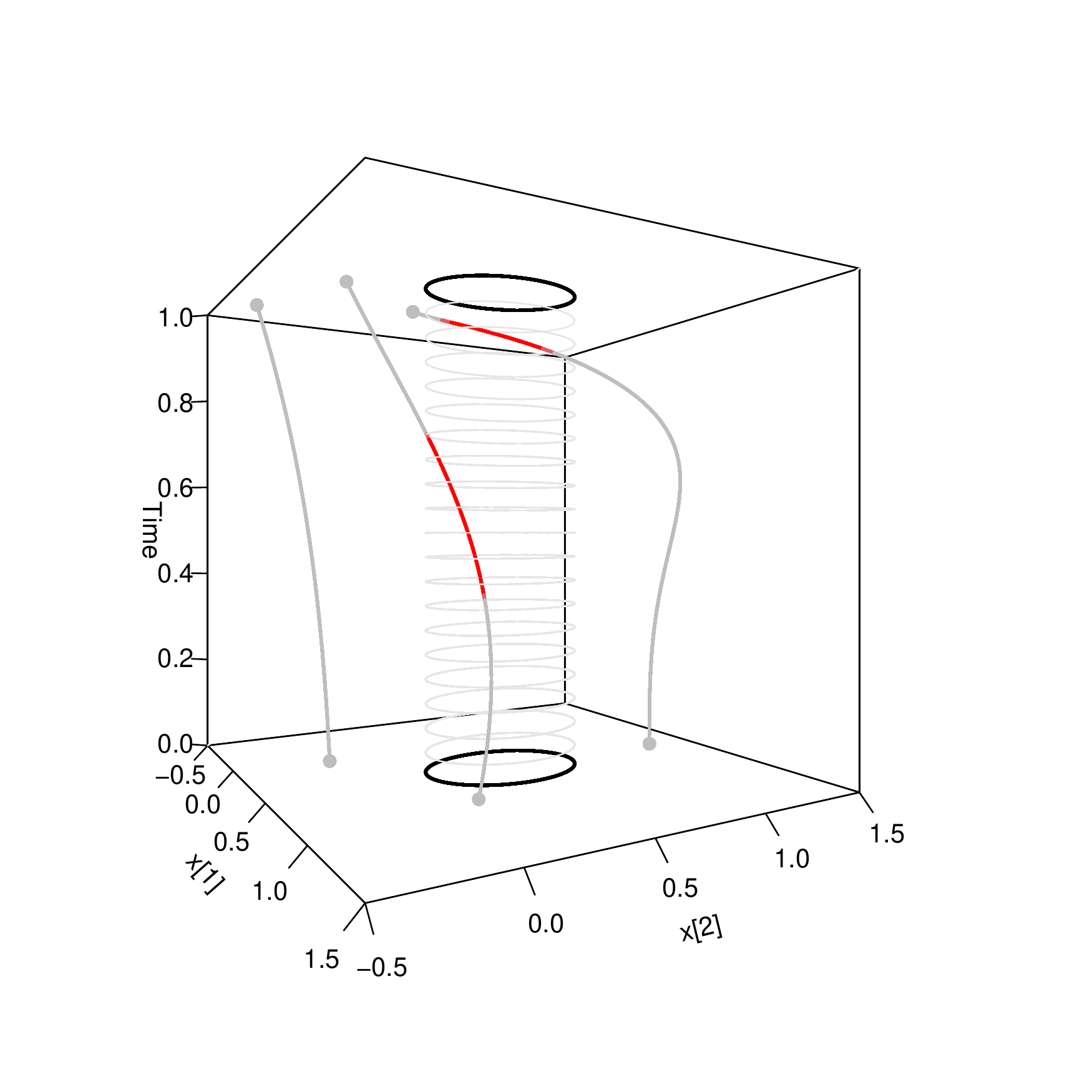} 
    \label{intersection_1}
        \vspace{-2\baselineskip}
    \caption{} 
  \end{subfigure}
  \begin{subfigure}[b]{0.5\linewidth}
    \centering
    \includegraphics[scale=0.5]{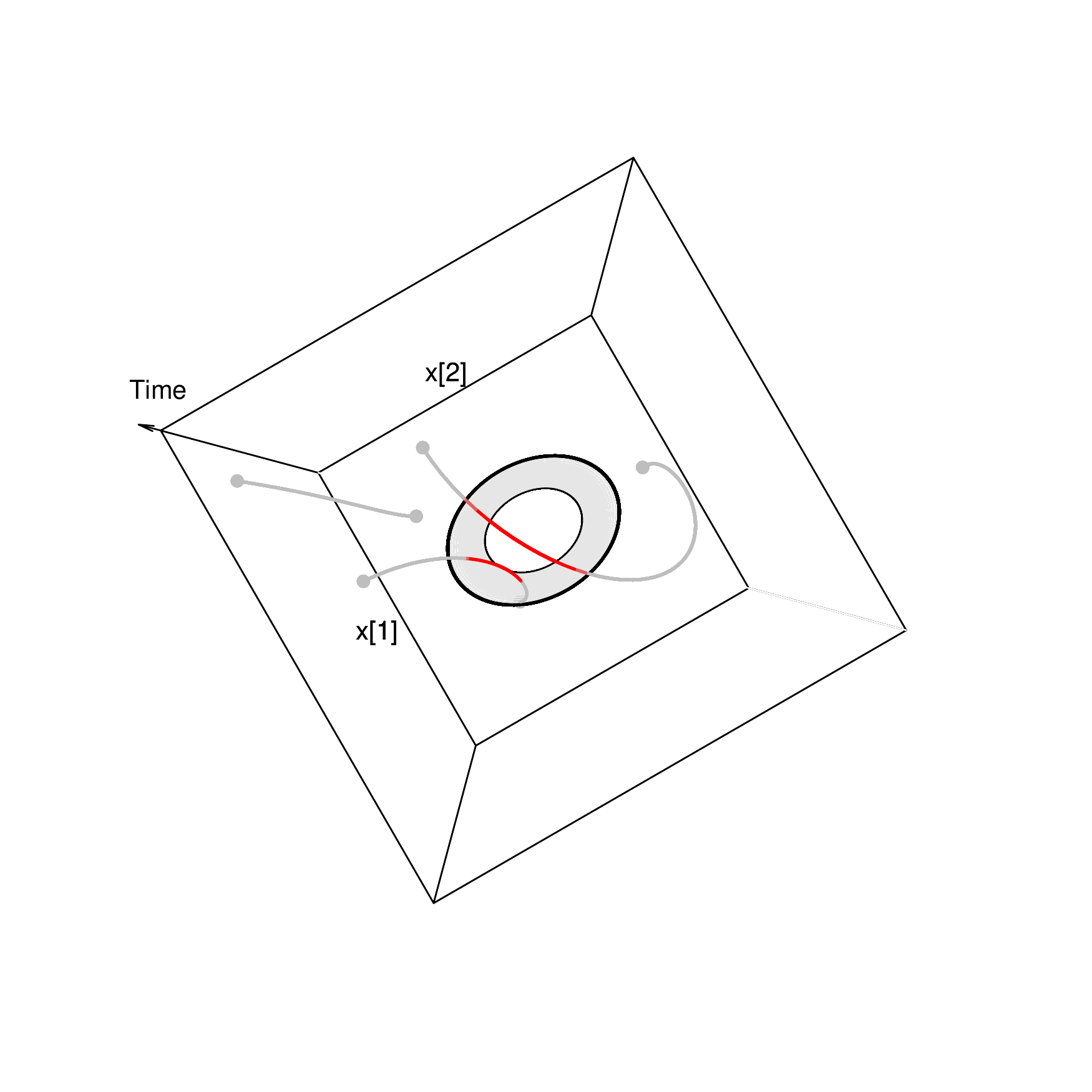}
      \label{intersection_2}
          \vspace{-2\baselineskip}
    \caption{} 
  \end{subfigure} 
  \caption{Illustration of contact-based, duration-based and length-based pointwise connectivities (resp. $\Psi_C$, $\Psi_D$ and $\Psi_L$) between the elliptic spatial domain $A\subset \mathbb R^2$ and different spatial points $x$ at time $s=1$, for $\Delta_{ts}=[0,1]$. The left curve on panel (a) never enters the domain $A$. The middle curve on panel (a) enters $A$ (red part of the curve) over a relatively long duration (as shown by panel (a)) but a short distance (as shown by panel (b)).  The right curve on panel (a) enters $A$ over a shorter duration but a longer distance. Thus, $\Psi_C(A \mid t,s,x)$, $\Psi_D(A \mid t,s,x)$ and $\Psi_L(A \mid t,s,x)$ are zero for the left curve; $\Psi_C(A \mid t,s,x)=1$ for the two other curves; $\Psi_D(A \mid t,s,x)$ is larger for the middle curve than for the right one, whereas $\Psi_L(A \mid t,s,x)$ is larger for the right curve than for the middle one.
  }
  \label{intersection}
  \end{figure}

Diverse types of the pointwise connectivity can be constructed, either using trajectory segments generated by $\Phi$, or directly using $\Phi$. Specific pointwise connectivities can include environmental covariates and even covariates associated to very the movements of particles. Below, we give several examples of such specifications. Some of these examples are graphically represented in Figure \ref{intersection}. Most examples are particularly relevant when $\Omega$ is a simple geographic domain and when $\Phi$ defines movements of {\it individuals} (e.g., air masses, animals or particles) within $\Omega$. 

\begin{exmp}\label{ex:contact}
 The {\it contact-based pointwise connectivity} is defined by:
    \begin{equation} 
    \label{eq:connect}
\Psi_{C}(A \mid t,s,x) = \mathds{1}_{\lbrace \mathcal{A}_{ts} \cap \Gamma(t,s,x) \neq \emptyset \rbrace},
    \end{equation}
    where $\mathcal{A}_{ts}= \Delta_{ts} \times A$ and $\mathds 1$ denotes the indicator function.
    $\Psi_C(A \mid t,s,x)$ indicates whether or not the particle whose movement in $\Omega$ is governed by $\Phi(\cdot, s, x)$ hit $A$ during the time interval $\Delta_{ts}$. Note that $A\mapsto\Psi_C(A\mid t,s,x)$ is only a submeasure on $\Omega$ since $\Psi_C(A\cup A'\mid t,s,x)\le \Psi_C(A\mid t,s,x)+\Psi_C(A'\mid t,s,x) $ for disjoint sets $A$ and $A'$ of $\mathcal B(\Omega)$.
 \end{exmp}


\begin{rem}
This example based on the simple contact between sets can be considered as too strict from a statistical and measure-theory perspective since the length or the duration of a contact may be null. Instead, a positive constraint on contact length for example can be used to define another version of the contact-based pointwise connectivity: Equation \eqref{eq:connect} could then be replaced by
    $$\Psi_{\tilde{C}}(A \mid t,s,x) = \mathds{1}_{\lbrace \mathcal L (\mathcal{A}_{ts} \cap \Gamma(t,s,x))>0 \rbrace}, $$
where $\mathcal L(\mathcal{A}_{ts} \cap \Gamma(t,s,x))$ denotes the length of the curve $\Gamma$ within $A$. The length operator $\mathcal L$ will be made explicit in Example \ref{ex:length}.

\end{rem}

  \begin{exmp}\label{ex:duration}
The {\it duration-based pointwise connectivity} is defined by:
   \begin{equation}
   \label{eq:dur}
   \Psi_{D}(A \mid t,s,x) = \int_{\Delta_{ts}} \mathds{1}_{\left\{ \Phi(v,s,x) \in A \right\}} dv,
   \end{equation}
to measure the duration spent by the particle in $A$ during $\Delta_{ts}$.
   \end{exmp}

\begin{exmp}\label{ex:length}
  The {\it length-based pointwise connectivity} is defined by:
   \begin{equation} 
   \label{eq:length} 
   \Psi_{L}(A \mid t,s,x) = \int_{\Delta_{ts}} \mathds{1}_{\left\{ \Phi(v,s,x)\in A \right\}} || \nabla_v \Phi (v,s,x) || dv ,
   \end{equation}
 where $\nabla_v \Phi (v,s,x)$ stands for the gradient of the flow $\Phi$ with respect to the time variable $v$ and $||\cdot||$ denotes the Euclidean norm.
  $\Psi_{L}(A \mid t,s,x)$ measures the distance travelled within $A$ by the particle during $\Delta_{ts}$.
  \end{exmp}

\begin{exmp}\label{ex:volume}
 The {\it pointwise connectivity based on local volume} is defined by:
    \begin{equation}
    \label{eq:volume}
    \Psi_{V}(A \mid t,s,x) = \int_{\Delta_{ts}} \mathds{1}_{\left( \Phi(v,s,x)\in A \right)} | \det(J^{x}_{\Phi}(v,s) | dv 
    \end{equation}
where $\det(J^{x}_{\Phi}(v,s))$ is the determinant of the Jacobian matrix (with respect to x) of the spatial transformation $\Phi (v,s,\cdot)$. The absolute value $| \det(J^{x}_{\Phi}(v,s) |$ of the Jacobian determinant at $x$ gives the ratio by which the function $\Phi (v,s,\cdot)$ expands/shrinks infinitesimal volumes around location $x$ into infinitesimal volumes around location $\Phi (v,s,x)$. 
In other words, $\Psi_{V}(A \mid t,s,x)$ assesses how particle density increases or decreases from $x$ to $A$ along the time interval $\Delta_{ts}$.
Intuitively, if $n$ particles are initially in $A$ and if the infinitesimal volume around any of these particles tends to shrink from $A$ to $x$, then one expects a high concentration of particles in a fixed volume around $x$ and, therefore, a high connectivity from $A$ to $x$. Conversely, if the infinitesimal volume around a particle tends to expand from $A$ to $x$, then one expects a lower concentration of particles in the same fixed volume around $x$ and, therefore, a lower connectivity from $A$ to $x$.
 \end{exmp}


More sophisticated specifications of the pointwise connectivity can be proposed by incorporating spatio-temporal covariates in its formulation, like in the following examples.

\begin{exmp} Let $G$ denote a time-varying vector field defined over $\mathbb R\times \Omega$. The {\it pointwise connectivity based on the external vector field $G$} is defined by:
    $$\Psi_{G}(A \mid t,s,x) = \int_{\Delta_{ts}} \mathds{1}_{\left( \Phi(v,s,x)\in A \right)} \ | < \nabla_v \Phi(v,s,x) ,G(v,\Phi(v,s,x))> | \ dv $$
where $< \nabla_v \Phi(v,s,x) ,G(v,\Phi(v,s,x))>$ is the scalar product between the gradient with respect to the time variable $v$ of the flow $\Phi$ and the vector field $G$. Larger the average collinearity in $A$ between the instantaneous movement of the particle and the simultaneous direction of the vector field $G$, higher the connectivity between $A$ and $x$. For instance, if $\Phi$ gives the movement of air masses and $G$ provides the intensity and the direction of a continuous release of specific particles, then the connectivity will be high (resp. low) if the movement of the air in $A$ and the movement of particles released in $A$ are approximately collinear (resp. orthogonal). 

\end{exmp}

\begin{figure}[H] 
\centering
  \begin{subfigure}[b]{0.5\linewidth}
    \centering
    \includegraphics[scale=0.5]{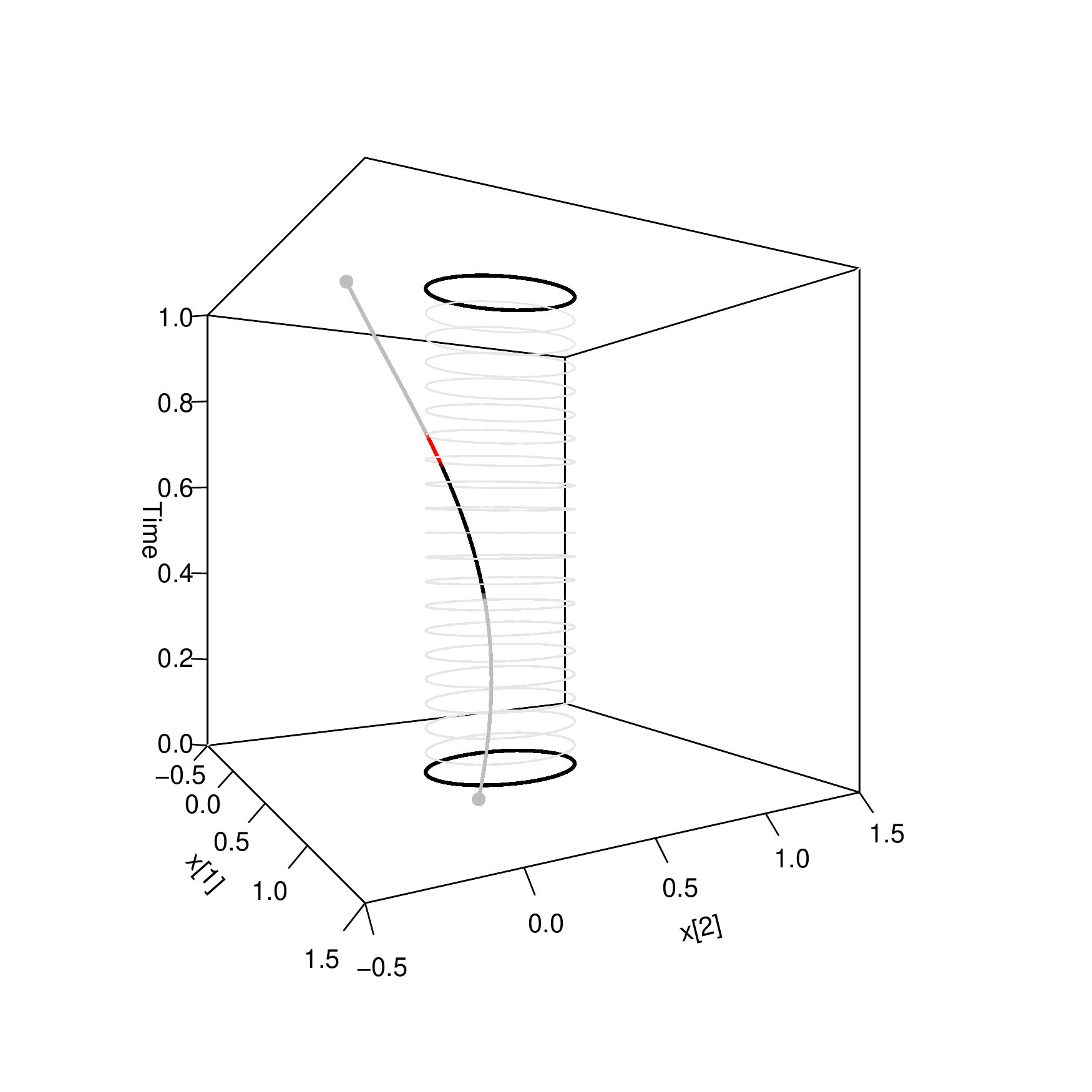} 
    \label{alt_intersection_1}
    \vspace{-2\baselineskip}
    \caption{} 
    \vspace{4ex}
  \end{subfigure}
  \begin{subfigure}[b]{0.5\linewidth}
    \centering
        \includegraphics[width=8cm,height=8cm]{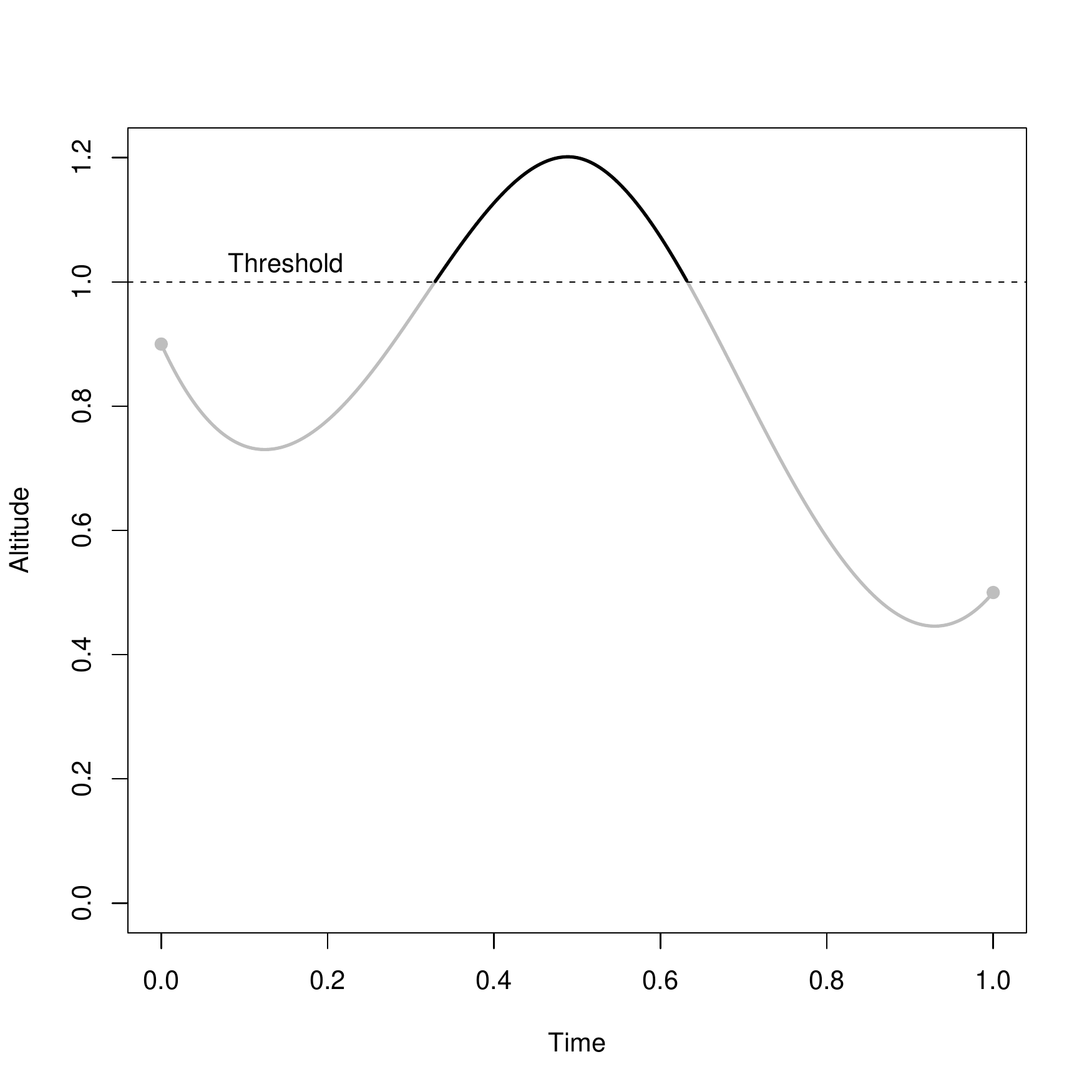}
       \vspace{-1\baselineskip}
   \caption{}
   \label{alt_intersection_2}
    \vspace{4ex}
  \end{subfigure} 
   \caption{Illustration of pointwise connectivity based on a covariate measured along the trajectory (see Example \ref{ex:covar}). In this illustration, the passage of the particle in the elliptic spatial domain $A$ contributes to the pointwise connectivity (red part of the curve in panel (a)) only when the particle is at an altitude lower than a threshold value (grey part of the curve in panel (b)).
   } 
     \label{alt_intersection}

  \end{figure}

 
 \begin{exmp}\label{ex:covar}
Let $Z$ and $\tilde Z$ be positive real valued spatio-temporal functions defined over $\mathbb R\times \Omega$. The pointwise connectivity based on $Z$ and $\tilde Z$ is defined by:
 \begin{equation}\label{eq:pC-Z}
      \Psi_{Z,\tilde Z}(A \mid t,s,x) = Z(s,x) \int_{\Delta_{ts}} \mathds{1}_{\left( \Phi(v,s,x)\in A \right)} \tilde Z(v, \Phi(v,s,x)) dv.
 \end{equation}

This form of pointwise connectivity may represent, for example, (i) the negative effect of the altitude of the air mass when it is above $A$ on the recruitment of specific particles from the ground, and (ii) the positive effect of rainfall at $(s,x)$ on the deposition of particles from the air mass to the ground (see Figure \ref{alt_intersection}). Thus, lower the average altitude of the air mass above $A$ and more intense the rainfall at $(s,x)$, larger the contribution to the connectivity between $A$ and $x$. This is expressed in Equation \eqref{eq:pC-Z} as follows: (i) $\tilde Z$ is defined as the binary function indicating whether or not the altitude of the air mass (located at $x$ at time $s$) is lower than a threshold $h$ when it is located at $\Phi(v,s,x)$ at time $v$; (ii) 
$Z$ is a function of the local rainfall intensity at $(s,x)$. 

 \end{exmp}

\begin{rem}
If in Example \ref{ex:covar}, the altitude of the air mass is incorporated as the third coordinate of $\Phi$ and $A$ is a 3D-domain vertically limited by the threshold value $h$, then, Equation \eqref{eq:pC-Z} is simplify reduced to Equation \eqref{eq:dur}. 
\end{rem}

 \begin{rem}
Example \ref{ex:covar} could be generalized by considering a measure, say $\mu$, over $\mathbb R$, to handle the potential contribution of discrete-time events to the pointwise connectivity:
\begin{equation}\label{eq:covar.measure}
  \Psi_{Z,\mu}(A \mid t,s,x) = Z(s,x) \int_{\Delta_{ts}} \mathds{1}_{\left( \Phi(v,s,x)\in A \right)} \tilde Z(v, \Phi(v,s,x)) d\mu(v). 
\end{equation}
\end{rem}

\begin{rem}
In the same vein, Example \ref{ex:covar} can also be modified by adding within the integral the term $|| \nabla_v \Phi (v,s,x) ||$ arising in Equation \eqref{eq:length} to account for a supplementary effect of the distance travelled within $A$ on the pointwise connectivity. 
\end{rem}

Each pointwise connectivity defined above can be used for defining the integrated connectivity, which measures the quantitative directional link between two subsets $A$ and $B$ of $\mathcal{B}(\Omega)$ generated by trajectories of particles located in $B$ at times belonging to the temporal domain $T$.


 \begin{definition}\label{def:Iconnect}
Let $A$ and $B$ be two sets of $\mathcal B(\Omega)$ and $T$ a subset of the temporal domain $\mathbb R$.
The {\it $\delta$-lag integrated connectivity} linking $B$ to $A$ over $T$ is defined by:
\begin{equation}
\label{eq:int_connectivity}
\Psi^{(2)}_{\nu, \delta} ( B \times A \mid T)= \int_{T \times B} 
\Psi(A \mid s+\delta,s,x) \nu (ds,dx),
\end{equation}
where $\delta\in\mathbb R$ and $\nu$ is a measure on $\mathbb{R} \times \Omega$.
\end{definition}



Definition \ref{def:Iconnect} encompasses  connectivities generated by either forward or backward trajectories, depending on the sign of $\delta$. The use of a unique duration $|\delta |$ could be relaxed to account for space-time heterogeneities in the duration of trajectories. It could even be infinite by introducing a measure over time like in Equation \eqref{eq:covar.measure}.

The measure $\nu$ in Definition \ref{def:Iconnect} can be continuous, discrete or hybrid over $\mathbb{R} \times \Omega$. Indeed, if particles of interest are air masses, then $B$ can be considered as continuously filled in space and time. Conversely, if particles of interest are animals of a specific species, then animals occupy only punctual locations in $B$ at each time and the measure $x\mapsto \nu(ds,dx)$, given $s$, is discrete in $\Omega$, whereas the temporal component of $\nu$ is continuous. Another examples occurs when the time $s$ corresponds to death times of animals, then $\nu$ is both discrete in space and time with a mass only at a countable collection of space-time points.

\subsection{Trajectory-based network}

\begin{definition}\label{def:TBN}
A {\it trajectory-based network} generated by $\Psi^{(2)}_{\nu, \delta}$ (given by Equation \eqref{eq:int_connectivity}) over the  temporal domain $T\subset\mathbb R$, is a graph whose nodes $A_i, i=\lbrace 1,..,I \rbrace$, are disjoint sets of $\Omega$ in $\mathcal B(\Omega)$ and whose directed edges are weighted by integrated connectivities 
$M_{ij}=\Psi^{(2)}_{\nu, \delta}( A_i \times A_j \mid T)$, $1\le i,j \le I$ and $i\neq j$.
\end{definition}


Definition \ref{def:TBN} corresponds to a spatial trajectory-based network evaluated over the fixed temporal domain $T$. It can be extended in different ways to obtain spatiotemporal analogs. For example, if $T_1,\ldots,T_K$ denote $K$ disjoint but successive time intervals with equal lengths, then the sequence of trajectory-based networks generated by $\Psi^{(2)}_{\nu, \delta}(\cdot \times \cdot \mid T_k)$, $k=1,\ldots,K$, forms a spatiotemporal trajectory-based network that can be analyzed to assess how connectivities across space are changing with time. This is one of the issues considered in Section \ref{subsec:net_analysis}.

\section{Estimation of integrated connectivities}\label{sec:estimation}

In practice, the integral defining the integrated connectivities between subsets of $\Omega$ (Definition \ref{def:Iconnect}) cannot be analytically computed in general, but can be estimated from a sample of trajectories. For instance, the integrated connectivity $\Psi^{(2)}_{\nu,\delta}( B \times A \mid T)$ can be estimated by its empirical counterpart obtained by importance sampling, say $\widehat{\Psi}^{(2)}_{\nu,\delta}( B \times A \mid T)$:
\begin{equation}
\label{estimation}
\widehat{\Psi}^{(2)}_{\nu,\delta}( B \times A \mid T)=  \dfrac{|T|}{N} \dfrac{|B|}{N'} \sum \limits_{k=1}^N \sum \limits_{l=1}^{N'} \Psi (A \mid s_k+\delta,s_k,b_l) ,
\end{equation}
where $s_1,\ldots ,s_{N} \in T$ and $b_1,\ldots ,b_{N'}\in B$ are times and locations, respectively, randomly drawn under the measure $\nu$ restricted to $T\times B$, $|T|$ and $|B|$ are the length and area of $T$ and $B$, respectively, and $\Psi (A \mid s_k+\delta,s_k,b_l)$ is the pointwise connectivity associated to the trajectory of the particle located at $b_l$ at time $s_k$ and observed over $\Delta_{s_k+\delta,s_k}=\left[ s_k \wedge s_k+\delta, s_k \vee s_k+\delta \right]$.

If $\nu$ is constant, other classical numerical approaches can be applied to approximate the integral, such as an hybrid approach in which the mid-point rule is applied in time and a regular point process is used in space. In such a case, the integrated connectivity estimator is also given by Equation \eqref{estimation}.

\begin{exmp}\label{ex:contact.est}
Using Equation \eqref{estimation}, the contact-based pointwise connectivity in Example \ref{ex:contact} is estimated by:
    \begin{equation} 
    \label{eq:connect.est}
\widehat\Psi^{(2)}_{C,\delta}(B\times A \mid T) = \dfrac{|T|}{N} \dfrac{|B|}{N'} \sum \limits_{k=1}^N \sum \limits_{l=1}^{N'} \mathds{1}_{\lbrace \mathcal{A}_{s_k+\delta,s_k} \cap \Gamma(s_k+\delta,s_k,b_l) \neq \emptyset \rbrace},
    \end{equation}
    where $\mathcal{A}_{s_k+\delta,s_k}= \Delta_{s_k+\delta,s_k} \times A$. Thus, $\widehat{\Psi}^{(2)}_{C,\delta} (B\times A \mid T) $ is simply the proportion of sampled trajectories intersecting $A$, multiplied by $|T||B|$.
 \end{exmp}

\begin{exmp}\label{ex:duration.est}
Using Equation \eqref{estimation}, the duration-based pointwise connectivity in Example \ref{ex:duration} is estimated by the average duration of the intersections between the sampled trajectories and $A$, multiplied by $|T||B|$.
\end{exmp}

\begin{exmp}\label{ex:length.est}
Using Equation \eqref{estimation}, the length-based pointwise connectivity in Example \ref{ex:length} is estimated by the average length of the intersections between the sampled trajectories and $A$, multiplied by $|T||B|$.
\end{exmp}

\section{Applications}\label{application}

In this section, we applied our general framework to the flow of air mass movements. Indeed, these movements compiled over years were used to characterize climatic patterns \citep{hondula2010back} and to describe the transport of pollutants \citep{perez2015applications}. We show now how to deploy our approach for constructing air-mass movement networks in two real geographical contexts, namely the coastline of the Mediterranean Sea and the French region of Provence-Alpes-C\^{o}te d'Azur. These two examples have been chosen in order to prove the flexibility of our approach to different situations and geographical scales.

\subsection{Case study regions and network construction} 

The first study region corresponds to the coast of the Mediterranean Sea, ranging approximately 1,600 km from north to south and  3,860 km from east to west. The temperate climate of the chosen region is strongly influenced by the presence of the Mediterranean Sea, with mild winters, hot summers and relatively scarce precipitations events. The landscape is characterized by coastal vegetation, typically shrubs and pines, and densely populated areas with intensive crop production of wheat, barley, vegetables and fruits, especially olive, grapes and citrus.
In this paper, we characterize recurrent movements of air masses through the Mediterranean region by defining a grid with mesh size 74 km covering the coastline from 5 km up to 250 km inland from the coast, including the four largest islands (namely Sicily, Sardinia, Cyprus and Corsica).
Thus, we divide the region into 604 cells, where the centroids of the cells will be used as arrival locations of air-mass trajectories and will correspond to the nodes of the constructed network.

The second study region corresponds to the French region of Provence-Alpes-C\^{o}te d'Azur (PACA, hereafter), located in the south-eastern part of France and characterized by a rather complex landscape formed by a densely-populated coastline, agricultural lands (high-value-crops with fruit and olive orchards, vineyards, vegetable cultivation and horticulture), and natural mostly-alpines areas. The choice of this particular region is justified in the context of a research project aimed at assessing the potential long-distance dissemination of phytopathogenic bacterial populations that are known to be transported by air currents. The bacteria of interest (e.g., {\it Pseudomonas syringae}) can be lifted in to the air from a source location and then be passively transported by air masses until they are deposited back to land onto a different, far away sink location. Since the life cycles of the considered species of bacteria are strongly linked to the water cycle \citep{morris2008life}, we naturally partitioned the study area in a way that fit this assumption. Hence, we considered the 294 watersheds of the PACA region to define the sites that will later constitute the nodes of the constructed network. Since watersheds have irregular shapes and varying sizes, we selected a certain number of arrival locations per watershed (between 1 and 10 and proportionally to the watershed area) in order to cover the watersheds consistently according to the relative importance of their size and estimate the integrated connectivities. In total, a set of 833 arrival locations for air-mass trajectories was generated. 

Once the arrival points for the two study regions have been established, we turned to the computation of air-mass trajectories arriving at the prescribed locations using the Hybrid Single-Particle Lagrangian Integrated Trajectory model (\verb"HYSPLIT", \citep{stein2015noaa}). The \verb"HYSPLIT" model can be fed with meteorological data from the Global Data Assimilation System files with a 0.5-degree spatial resolution (GDAS\footnote{\url{https://www.ncdc.noaa.gov/data-access/model-data/model-datasets/global-data-assimilation-system-gdas}}) and was tuned by us to return 48-hours backward air-mass trajectories arriving at the prescribed locations at an altitude of 500 m above mean sea level. A single trajectory consists of a vector containing the hourly positions (longitude, latitude and altitude) visited by the air mass before arriving at the specified location and time.
Air-mass trajectories have been computed for every arrival location (604 for the Mediterranean region and 833 for the PACA region) and for every day between January 1, 2011 and December 31, 2017 (arrival hour is 12:00 GMT). The total number of computed trajectories is 1,543,220 for the Mediterranean region and 2,128,315 for the PACA region. 

The final step for the construction of the networks is the estimation of the adjacency matrices of the networks, based on the methodology presented in the previous sections. To do that, for each pair of subsets of the spatial domain, we used the daily 48-hours backward trajectories arriving at the locations sampled within the receptor subset, and computed the contact-based estimator (see Example~\ref{ex:contact.est}). The subsets of the spatial domain are the watersheds for PACA and circular buffers of radius 20 km for the Mediterranean region, as in \cite{leyronas2018assessing}

In this work we will consider networks corresponding to three temporal contexts: (i) the spatial networks obtained when $T$ is the entire period 2011-2017, (ii) the yearly spatiotemporal networks formed by the seven spatial networks obtained when $T_1$ encompasses the year 2011, $T_2$ encompasses 2012 and so on, and (iii) monthly spatiotemporal networks formed by the twelve spatial networks obtained when $T_1$ represents every January from 2011 to 2017, $T_2$ every February from 2011 to 2017, and so on. In all these cases, we consider that the length of the time interval was 1 to easily compare the inferred networks (i.e., $|T|=1$ in Equations \eqref{estimation} and \eqref{eq:connect.est}).

\subsection{Network analysis} \label{subsec:net_analysis}

The constructed networks are directed and weighted by contact-based connectivities generated by air mass trajectories. They are inherently complex by the sheer amount of spatial and temporal information that they encompass. Hence, there is no easy way of representing the results either graphically or numerically, without compromising the original complexity of the networks. While a comprehensive physical study of the spatiotemporal properties of these networks goes beyond the scope of the paper, we explore the estimated trajectory-based networks by looking at some generic properties through the following indices:
\begin{itemize}
\item Diameter: the longest of all possible shortest paths between any two pair of nodes.
\item Density: the ratio between the sum of all edge weights and the number of all possible edges \citep{liu2009complex}.
\item Transitivity (also known as clustering): the equivalent definition of density, but applied to triplets of nodes instead of pairs of nodes \citep{opsahl2009clustering}.
\item Shortest path: characterized by the average and standard deviation of the computed shortest path between any possible pair of different nodes \citep{newman2001scientific}.
\item Small worldness: refers to the property of a network of being highly clustered and having relatively short shortest paths. It is computed as the ratio between the normalized clustering and the normalized average shortest path distance \citep{li2007structure, colon2016small}.
\item Scale-free property: The degree of a node in terms of the total number of edges entering and exiting from it, and for directed networks it can be decomposed in the incoming and outgoing degree, respectively. The degree distribution is the empirical distribution of the degree of a network and it said to be scale free when it approximately follows a power law distribution, i.e. $P(k) \sim k^{(-\alpha)}$, where $P(k)$ represents the probability of a node having degree equal to $k$ \citep{barabasi2003scale,barabasi1999emergence}. Some authors impose that the $\alpha$ parameter of the power law distribution has to fall within the interval $[2,3]$ \citep{barabasi2016network}. Thus, a network is scale free when most of its nodes have low degree, while the probability of having nodes with very high degree is not negligible (fat right tail of the distribution). Nodes with very high degree play a crucial role in dynamics conditional on networks and are often referred as hubs \citep{liu2011controllability}.
\item Degree correlation: in directed networks, it accounts for the correlation between the incoming and the outgoing degree of a node. Networks with positive (resp. negative) degree correlation foster (resp. hamper) epidemic spread \citep{pautasso2010disease}.
\end{itemize}

\subsection{Results}\label{sec:results}

\begin{figure}[H] 
  \begin{subfigure}[b]{0.5\linewidth}
    \centering
    \includegraphics[scale=0.3]{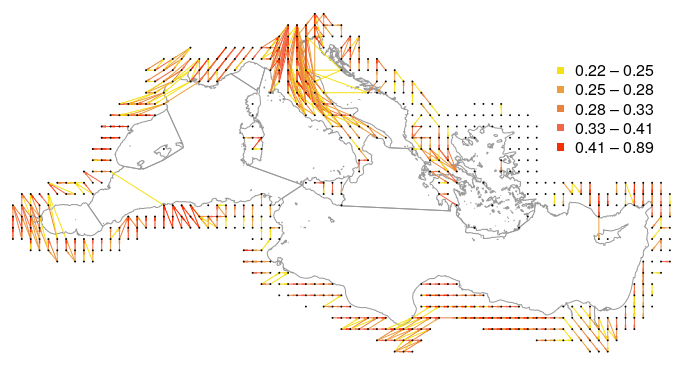} 
    \caption{} 
    \label{med_network} 
  \end{subfigure}
  \begin{subfigure}[b]{0.5\linewidth}
    \centering
    \includegraphics[scale=0.3]{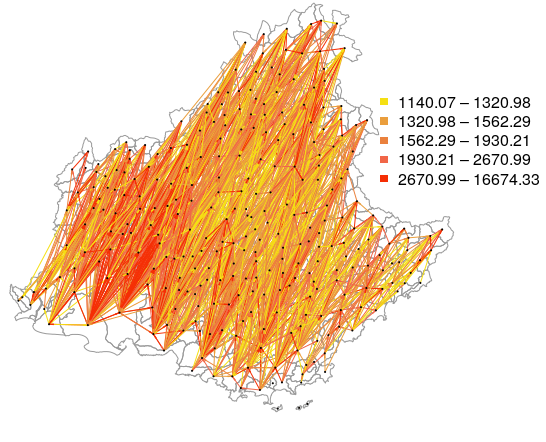}
   \caption{}
\label{paca_network}
  \end{subfigure} 
 \caption{Networks weighted by contact-based connectivities generated by air mass trajectories between (a) the 604 sampled circular areas within the Mediterranean basin and (b) the 294 watersheds of PACA. Edges with weights lower than 0.3 for (a) and $2 \times 10^3$ for (b) are not drawn. The cuts of the intervals in the two legends are chosen in such a way that each interval contains 20\% of the observed data. The differences in the values taken by the connectivities in (a) and (b) are due to different measures of the area $|B|$ in  Equations \eqref{estimation} and \eqref{eq:connect.est}: $|B|=1$ for each node in (a), whereas $B$ is the actual area (expressed in km$^2$) of each whatershed in (b).}
\label{network}
\end{figure}

The two spatial trajectory-based networks representing the strength of tropospheric connections in the Mediterranean region and PACA during the entire period 2011 to 2017 are represented in Figure \ref{network}. In order to highlight the edges that represent strong connections, we depicted them with darker shades of color, while we did not draw the connections that had a weight of less than 0.3 for the Mediterranean and $2 \times 10^3$ for PACA. It can be seen that the strongest connections tend to link nodes that are geographically close, but nonetheless moderate connections also exist between rather distant nodes (see also Figure \ref{Boxplot_dist}). This is confirmed by small values of the average shortest path distance $8.20 \times 10^{-4}$ for the Mediterranean and $2.57 \times 10^{-4}$ for PACA,
and high values of the transitivity index (0.74 for the Mediterranean region and 0.99 for PACA), as shown in the first lines of Tables \ref{tab:result1} and \ref{tab:result2}. The connectivities in PACA are mostly oriented from North-East to South-West, which corresponds to the direction of the prevailing wind in this region. For the Mediterranean basin, the direction of the connectivites depends on the region and does not have a fixed direction (see Figure \ref{Circular_ang}). An interesting additional difference between the two networks is that the one for PACA has a very negative degree correlation ($-0.85$), meaning that nodes having a high incoming degree will have low outgoing degree, and vice versa. On the other hand, for the Mediterranean network, the value of the degree correlation is moderately positive (0.31), meaning that nodes having high a incoming degree tend to have also high a outgoing degree.

\begin{table}[H]
{\renewcommand{\arraystretch}{1} 
{\setlength{\tabcolsep}{0.2cm} 
\begin{tabular}{|l|r|r|r|r|r|r|r|r|r|r|r|}
\hline
\multicolumn{9}{|c|}{\textbf{Mediterranean region}}  \\ \hline  
 & Diam & Dens & Trans & S P (mean) &  S P (sd)  & S W & S F & D C \\ \Xhline{3\arrayrulewidth}
 2011-2017 & 3.12$\times 10^{-3}$ & 2.96$\times 10^{-4}$ & 0.74 & 8.20$\times 10^{-4}$ & 2.19$\times 10^{-4}$ & 18242 & 11.6 & 0.31 \\ \Xhline{3\arrayrulewidth}
2011 & 0.03 & 3.0$\times 10^{-4}$ & 0.67 & 6.17$\times 10^{-3}$ & 1.92$\times 10^{-3}$ & 2222 & 11.5 & -0.04\\ \hline
  2012 & 0.02 & 2.9$\times 10^{-4}$ & 0.67 & 6.09$\times 10^{-3}$ & 1.87$\times 10^{-3}$ & 2236 & 10.4 & -0.06\\ \hline
  2013 & 0.01 & 3.0$\times 10^{-4}$ & 0.66 & 6.06$\times 10^{-3}$ & 1.86$\times 10^{-3}$ & 2228 & 13.9 & 0.20  \\ \hline
  2014 & 0.08 & 2.9$\times 10^{-4}$ & 0.65 & 6.52$\times 10^{-3}$ & 2.11$\times 10^{-3}$ & 2044 & 9.0 & 0.30 \\ \hline
  2015 & 0.06 & 3.0$\times 10^{-4}$ & 0.66 & 6.22$\times 10^{-3}$ & 1.93$\times 10^{-3}$ & 2147 & 8.2 & 0.26 \\ \hline
  2016 & 0.55 & 3.0$\times 10^{-4}$ & 0.66 & 6.32$\times 10^{-3}$ & 1.97$\times 10^{-3}$ & 2110 & 8.8 & 0.25\\ \hline
  2017 & 0.12 & 3.0$\times 10^{-4}$ & 0.65 & 6.08$\times 10^{-3}$ & 1.83$\times 10^{-3}$ & 2186 & 9.2 & 0.12  \\ \Xhline{3\arrayrulewidth}

January & 0.23 & 3.1$\times 10^{-4}$ & 0.63 & 1.15$\times 10^{-2}$ & 3.81$\times 10^{-3}$ & 1118 & 14.9 & 0.14 \\ \hline
  February & 0.67 & 3.0$\times 10^{-4}$ & 0.63 & 1.15$\times 10^{-2}$ & 3.92$\times 10^{-3}$ & 1106 & 14.9 & 0.24 \\ \hline
  March & 0.29 & 3.0$\times 10^{-4}$ & 0.62 & 1.15$\times 10^{-2}$ & 3.87$\times 10^{-3}$ & 1104& 11.9 & 0.26 \\ \hline
  April & 0.70 & 3.1$\times 10^{-4}$ & 0.64 & 1.13$\times 10^{-2}$ & 3.91$\times 10^{-3}$ & 1137 & 15.4 & 0.18  \\ \hline
  May & 1.02 & 3.2$\times 10^{-4}$ & 0.63 & 1.20$\times 10^{-2}$ & 4.43$\times 10^{-3}$ & 1064 & 12.4 & 0.21  \\ \hline
  June & 1.14 & 3.2$\times 10^{-4}$ & 0.61 & 1.19$\times 10^{-2}$ & 4.18$\times 10^{-3}$ & 1041 & 12.9 & 0.15  \\ \hline
  July & 1.41 & 3.1$\times 10^{-4}$ & 0.60 & 1.19$\times 10^{-2}$ & 4.01$\times 10^{-3}$ & 1017 & 28.2 & -0.07  \\ \hline
  August & 1.87 & 2.9$\times 10^{-4}$ & 0.59 & 1.22$\times 10^{-2}$ & 4.14$\times 10^{-3}$ & 987 & 11.9 & -0.02  \\ \hline
  September & 1.51 & 2.9$\times 10^{-4}$ & 0.60 & 1.23$\times 10^{-2}$ & 4.18$\times 10^{-3}$ & 994 & 12.8 & 0.13  \\ \hline
  October & 0.12 & 2.8$\times 10^{-4}$ & 0.62 & 1.26$\times 10^{-2}$ & 4.46$\times 10^{-3}$ & 998 & 14.9 & 0.12 \\ \hline
  November & 0.10 & 2.7$\times 10^{-4}$ & 0.62 & 1.27$\times 10^{-2}$ & 4.90$\times 10^{-3}$ & 997 & 14.3 & 0.41 \\ \hline
  December & 1.07 & 3.0$\times 10^{-4}$ & 0.61 & 1.16$\times 10^{-2}$ & 3.89$\times 10^{-3}$ & 1079 & 14.1 & 0.32 \\ \Xhline{3\arrayrulewidth}
\end{tabular}}}
\caption{Network indices (Diameter, density, transitivity, shortest path (mean and standard deviation), small worldness, scale-free property, degree correlation) calculated from the networks covering the Mediterranean region and estimated in three temporal contexts: the entire period 2011-2017, yearly time periods from 2011 to 2017 and monthly time periods.}
\label{tab:result1}
\end{table}

\begin{table}[H]
{\renewcommand{\arraystretch}{1} 
{\setlength{\tabcolsep}{0.2cm} 
\begin{tabular}{|l|r|r|r|r|r|r|r|r|r|r|r|}
\hline
\multicolumn{9}{|c|}{\textbf{PACA}}  \\ \hline  
 & Diam & Dens & Trans & S P (mean) &  S P (sd)  & S W & S F & D C \\ \Xhline{3\arrayrulewidth}
 2011-2017 & 6.06$\times 10^{-3}$& 2.51$\times 10^{-3}$ & 0.99  & 2.57$\times 10^{-4}$&  1.00$ \times 10^{-4}$ & 7813 & 19.2 & -0.85\\ \hline \Xhline{3\arrayrulewidth}
   2012 & $2.22 \times 10^{-3}$ & $9.9\times 10^{-4}$ & 0.97 & $6\times 10^{-4}$ & $1.7\times 10^{-4}$ & 3287.37 & 21.50 & -0.88 \\ \hline
  2013 & $2.42\times 10^{-3}$ & $1.01\times 10^{-3}$ & 0.98 & $6.5\times 10^{-4}$ & $1.8\times 10^{-4}$ & 3069.76 & 23.71 & -0.91  \\ \hline
  2014 & $4.23\times 10^{-3}$ & $ 10^{-3}$ & 0.98 & $7.3\times 10^{-4}$ & $2.2\times 10^{-4}$ & 2717.58 & 24.56 & -0.92  \\ \hline
  2015 & $2.92\times 10^{-3}$ & $1.02\times 10^{-3}$ & 0.99 & $7.2\times 10^{-4}$ & $2.2\times 10^{-4}$ & 2789.31 & 27.09 & -0.92 \\ \hline
  2016 & $4.18\times 10^{-3}$ & $1.01\times 10^{-3}$ & 0.98 & $8\times 10^{-4}$ & $2.5\times 10^{-4}$ & 2470.12 & 19.29 & -0.92  \\ \hline
  2017 & $2.46\times 10^{-3}$ & $1.01\times 10^{-3}$ & 0.98 & $5.9\times 10^{-4}$ & $1.6\times 10^{-4}$ & 3348.41 & 17.51 & -0.89  \\ \Xhline{3\arrayrulewidth}
January & 2.76 & $1.8\times 10^{-3}$ & 0.69 & $3.03\times 10^{-3}$ & $9.2\times 10^{-4}$ & 462.54 & 6.84 & -0.66  \\  \hline
  February & 4.13 & $1.64\times 10^{-3}$ & 0.73 & $3.35\times 10^{-3}$ & $1.09\times 10^{-3}$ & 439.54 & 8.23 & -0.6\\  \hline 
  March & 0.57 & $1.71\times 10^{-3}$ & 0.7 & $2.18\times 10^{-3}$ & $6.9\times 10^{-4}$ & 651.63 & 22.3 & -0.84  \\  \hline
  April & $1.34\times 10^{-2}$ & $1.69\times 10^{-3}$ & 0.99 & $2.81\times 10^{-4}$ & $8.8\times 10^{-4}$ & 712.56 & 43.44 & -0.98  \\  \hline
  May & 0.37 & $1.86\times 10^{-3}$ & 0.84 & $1.73\times 10^{-3}$ & $5.7\times 10^{-4}$ & 984.96 & 46.82 & -0.93  \\  \hline
  June & 5.05 & $1.79\times 10^{-3}$ & 0.7 & $5.28\times 10^{-3}$ & $2.04\times 10^{-3}$ & 270.61 & 6.96 & -0.64 \\  \hline
  July & 5.02 & $1.84\times 10^{-3}$ & 0.71 & $6.85\times 10^{-3}$ & $2.58\times 10^{-3}$ & 211.02 & 7.92 & -0.64\\  \hline
  August & 5.1 & $1.84\times 10^{-3}$ & 0.71 & $5.25\times 10^{-3}$ & $2.21\times 10^{-3}$ & 274.57 & 6.48 & -0.66 \\  \hline
  September & 5.1 & $1.84\times 10^{-3}$ & 0.64 & $2.57\times 10^{-3}$ & $7.6\times 10^{-4}$ & 505.03 & 6.46 & -0.73 \\  \hline 
  October & $1.11\times 10^{-2}$ & $1.87 \times 10^{-3}$ & 0.99 & $1.83\times 10^{-3}$ & $5.3\times 10^{-4}$ & 1101.63 & 25.46 & -0.92  \\  \hline
  November & $8.23\times 10^{-3}$ & $1.72\times 10^{-3}$ & 0.95 & $1.68\times 10^{-3}$ & $5.1 \times 10^{-4}$& 1147.04 & 29.41 & -0.89 \\  \hline
  December & 1.55 & $1.8\times 10^{-3}$ & 0.73 & $2.69\times 10^{-3}$ & $8.1\times 10^{-4}$ & 548.04 & 7.08 & -0.7 \\ \Xhline{3\arrayrulewidth}

\end{tabular}}}
\caption{Network indices (Diameter, density, transitivity, shortest path (mean and standard deviation), small worldness, scale-free property, degree correlation) calculated from the networks covering PACA and estimated in three temporal contexts: the entire period 2011-2017, yearly time periods from 2011 to 2017 and monthly time periods.}
\label{tab:result2}
\end{table}

Qualitatively, the indices provided in Tables \ref{tab:result1} and \ref{tab:result2} are overall more variable for the monthly spatiotemporal trajectory-based networks than for the yearly ones. Thus, focusing in what follows on the monthly networks, we investigate possible seasonal patterns by using the complete-linkage hierarchical clustering method \citep{ferreira2009comparison}. We applied the clustering using the Euclidean distance over the 8-dimensional space formed by the 8 indices provided in Tables \ref{tab:result1} and \ref{tab:result2}.

\begin{figure}
\begin{subfigure}{1\linewidth}
\centering
    \includegraphics[scale=0.5]{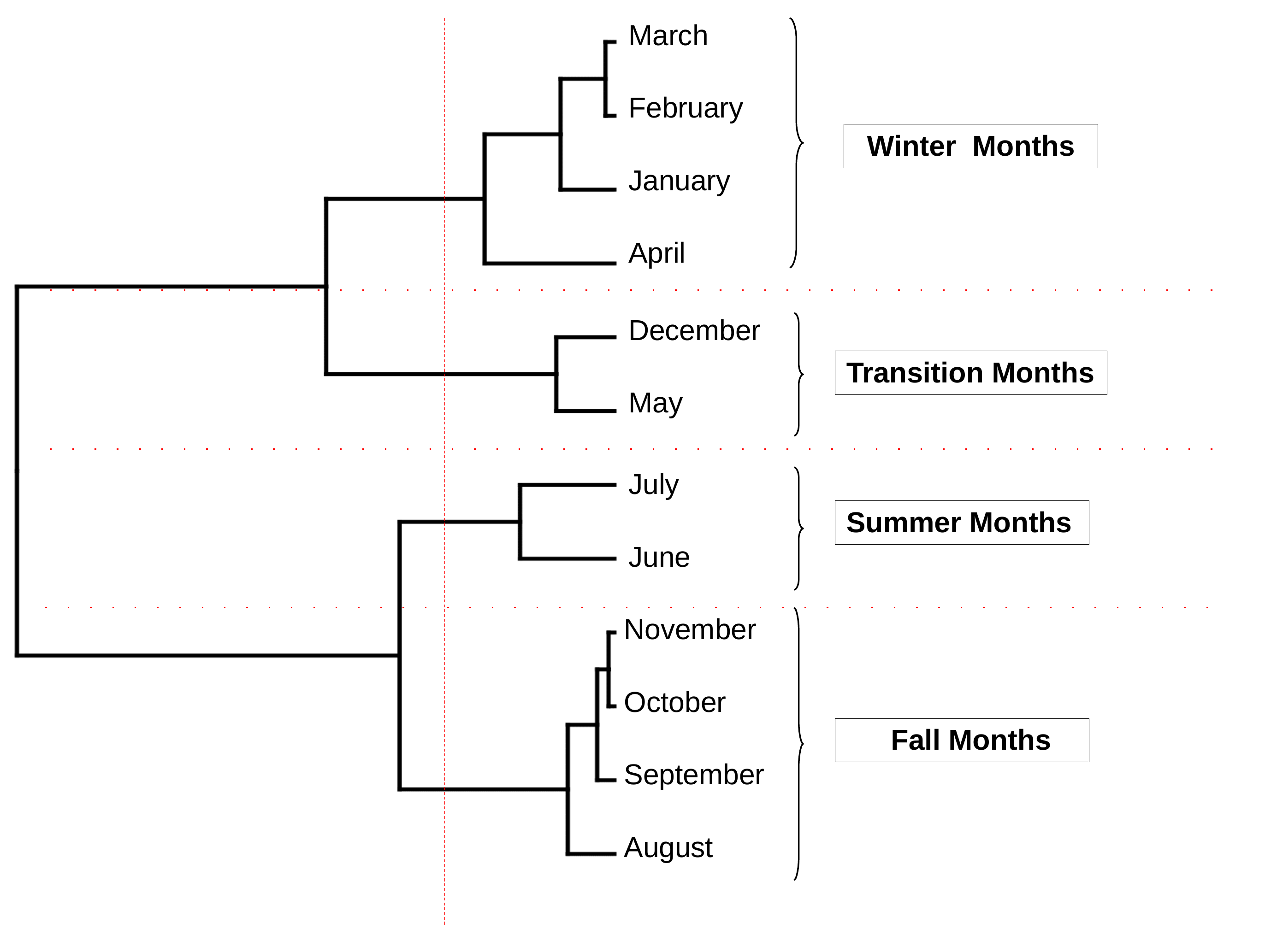}
    \caption{}
    \label{dendogram_med}
\end{subfigure}\\[1ex]
\begin{subfigure}{0.5\linewidth}
\centering
\includegraphics[scale=0.25]{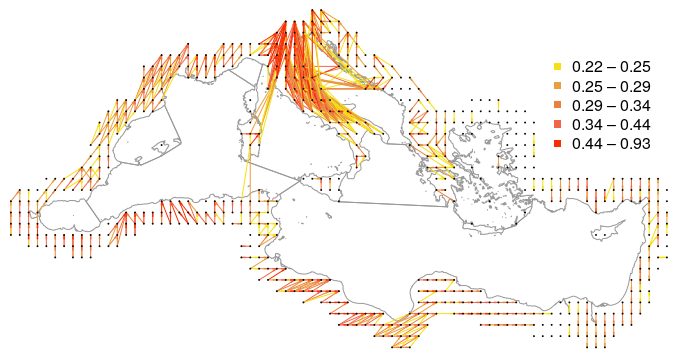} \caption{Winter Months}
\label{cl_1_med}
\end{subfigure}
\begin{subfigure}{.5\linewidth}
\centering
 \includegraphics[scale=0.25]{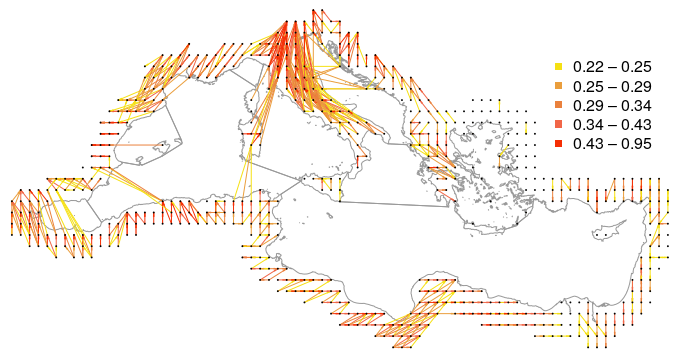}\caption{Transition Months}
\label{cl_2_med}
\end{subfigure}\\[1ex]
\begin{subfigure}{.5\linewidth}
\centering
 \includegraphics[scale=0.25]{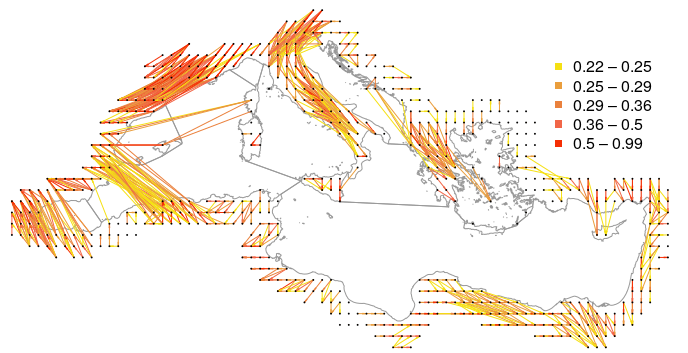}
 \caption{Summer Months}
\label{cl_3_med}
\end{subfigure}%
\begin{subfigure}{.5\linewidth}
\centering
 \includegraphics[scale=0.25]{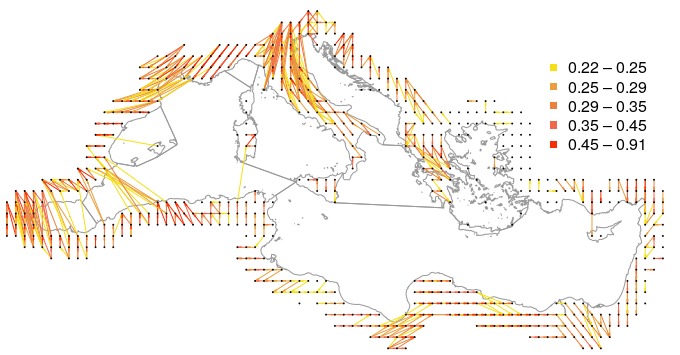}
 \caption{Fall months}
\label{cl_4_med}
\end{subfigure}%
\caption{(a): Dendrogram of the months obtained from a hierarchical cluster analysis of the Mediterranean spatio-temporal network based on the monthly dissimilarities of the indices presented in Table \ref{tab:result1}. (b), (c), (d) and (e): Networks corresponding to the four identified clusters where one displayed only the edges between the nodes connected more than 10 days per month via the air mass trajectories. }
\label{dend_Med}
\end{figure}

For the Mediterranean region, the dendogram in Figure \ref{dend_Med}(a) can be used to identify four distinctive periods: summer months (June and July), winter months (January, February, March, April), fall months (August, September, October, November)  and a set of transition months (May and December) surrounding the winter months. The spatial networks derived from this clustering are shown in Figure \ref{dend_Med}(b-e), which displays clear differences in the connectivity patterns even if one observes similarities between the networks for winter months and the surrounding transition months (winter and transition months are precisely in the same dendogram cluster if one increases the cut-off). The main differences are observed in the northwestern part of the Mediterranean basin with, in particular, increased connectivities in the North of Italy in Winter, in the South of France and the East of Spain in Summer, between Spain and Algeria / Morocco in Summer, and along the eastern Mediterranean coast in Summer. 

\begin{figure}[H]
    \begin{subfigure}[c]{1\linewidth}
        \centering
    \includegraphics[scale=0.6]{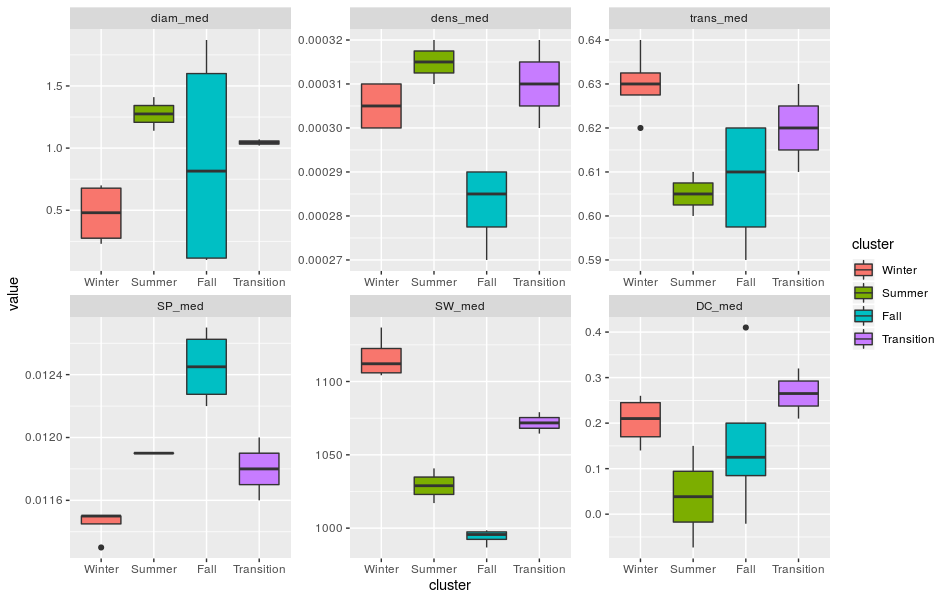}
            \caption{Mediterranean region}
\label{boxplot_Med}
    \end{subfigure}\\[1ex]
    \begin{subfigure}[c]{1\linewidth}
        \centering
    \includegraphics[scale=0.6]{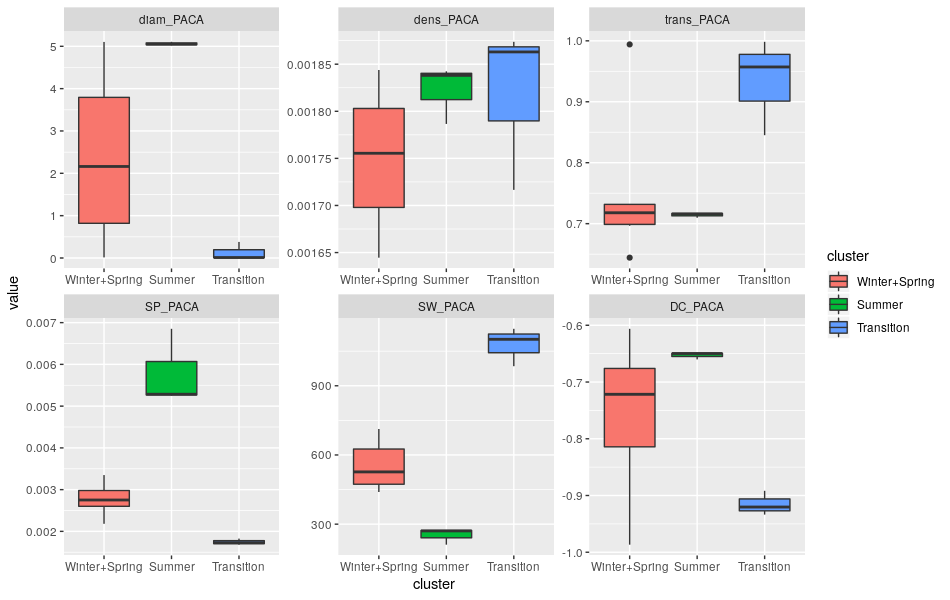}
                \caption{PACA}
        \label{boxplot_PACA}
    \end{subfigure}
    \caption{Boxplot for the different indices (Diameter, density, transitivity, shortest path (mean), small worldness, degree correlation) obtained from (a) the four clusters identified for the Mediterranean region (see Figure \ref{dend_Med}) and (b) the three clusters for PACA (see Figure \ref{dend_PACA}). }
\label{Boxplot}
\end{figure}

As shown by Figure \ref{Boxplot}(a), summer months are characterized by high diameter and density, low values of transitivity, small-worldness and the lowest values of degree correlation. 
Winter months have lower diameter and show high values of small-worldness due to its high values of clustering and low values of average shortest path distances. 
Fall months have lower values of density and small-worldness due to its low values of clustering and high values of average shortest path distances. 
Finally, the group of transition months show high values of density and degree correlation.

\begin{figure}

\begin{subfigure}{1\linewidth}
\centering
\includegraphics[scale=0.5]{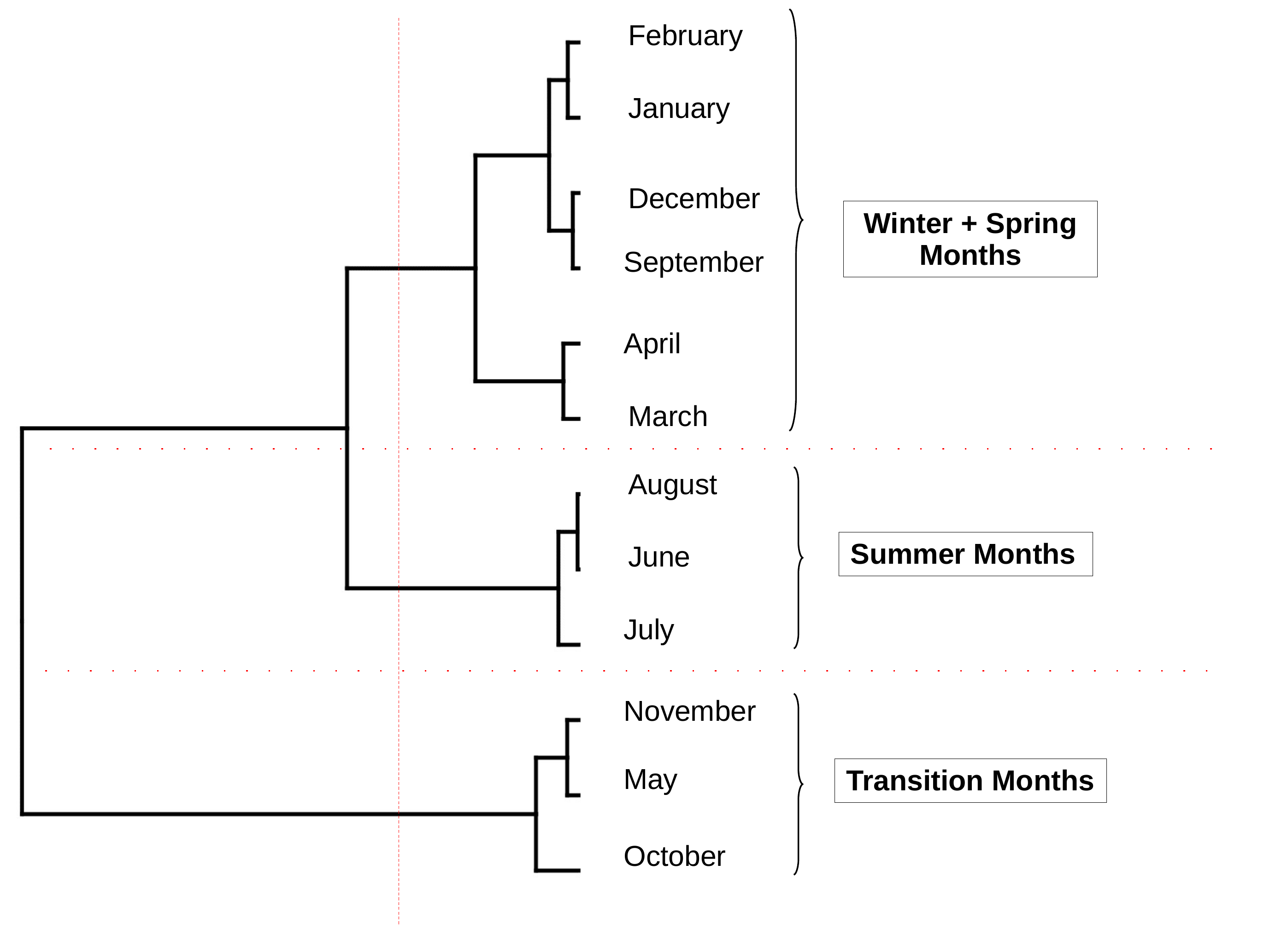}
\caption{}
\label{dendogram_paca}
\end{subfigure}\\[1ex]
\begin{subfigure}{0.3\linewidth}
\centering
\includegraphics[scale=0.25]{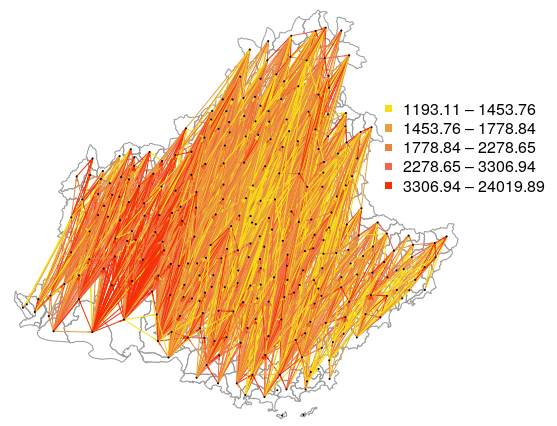} \caption{Winter + Spring months}
\label{cl_1_paca}
\end{subfigure}
\begin{subfigure}{.3\linewidth}
\centering
 \includegraphics[scale=0.25]{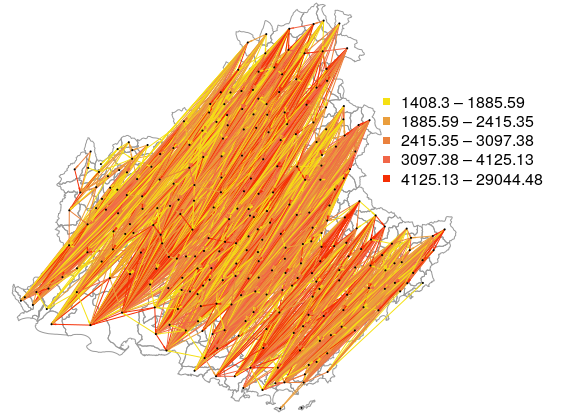}\caption{Summer months}
\label{cl_2_paca}
\end{subfigure}%
\begin{subfigure}{.3\linewidth}
\centering
 \includegraphics[scale=0.25]{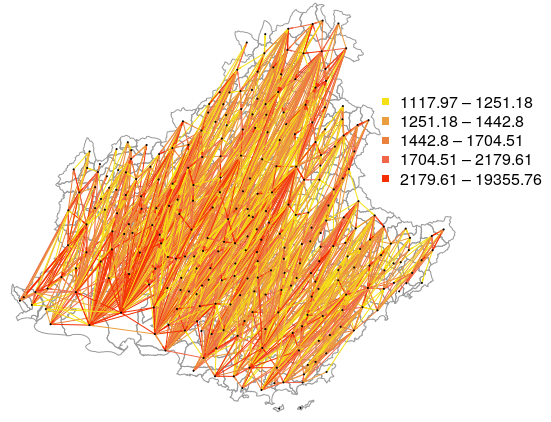}\caption{Transition months}
\label{cl_3_paca}
\end{subfigure}%
\caption{(a): Dendrogram of the months obtained from a hierarchical cluster analysis of the PACA spatio-temporal network based on the monthly dissimilarities of the indices presented in Table \ref{tab:result2}. (b), (c) and (d): Networks corresponding to the three identified clusters where one displayed only the edges between the nodes connected more than 10 days per month via the air mass trajectories. }
\label{dend_PACA}
\end{figure}

For PACA, the dendogram in Figure \ref{dend_PACA}(a) can be used to identify three distinctive periods: summer months (from June to August), winter and spring months (from December to April, plus September that can be considered as an outlier from a chronological viewpoint) and a set of transition months between the two previous periods (May, October and November). 

Figure \ref{dend_PACA}(b-d) illustrates the differences between the networks derived from this clustering. The summer network is largely more connected than the two other networks, and the transition months, surprisingly, do not lead to intermediate connectivities but to the lowest connectivities. Based on Figure \ref{Boxplot}(b),
the group of summer months is characterized by a high diameter, density and average shortest path, low values of transitivity, small-worldness and the lowest degree correlations (yet still negative). Winter and spring months have significantly lower diameter, density and average shortest path distances. Finally, the group of transition months show the highest values of small-worldness, due to their high values of clustering and low values of average shortest path distances.

\section{Discussion}
\label{sec:discussion}

We presented a framework for estimating and characterizing spatial and spatio-temporal networks generated by trajectory data. The development of this framework was motivated by the study of networks resulting from the movement of air masses sampled over long time periods and large spatial scales. Thus, in the application, we investigated the tropospheric connectivities across the Mediterranean basin and the French region PACA, and their variations through years and months. 
Our approach could be applied to diverse phenomena, from which trajectories can be observed. For instance, one could estimate networks generated by the movement of animals on the landscape scale based on animal trajectories observed with GPS devices \citep{bastille2018applying}. This would allow the characterization of connectivity between different landscape components. Sampled trajectories of humans, sampled transports of specific goods (such as plant material) and sampled trajectories of knowledge in social communities (that cannot be exhaustively observed) could also be used to estimate networks in other applied settings. 

In Section \ref{sec:pic}, we proposed diverse measures of connectivity with different underlying (physical or biological) interpretations. Thus, the analyst can adapt the connectivity measure to the mechanistic processes he investigates. In the application section, we only used the contact-based connectivity. Comparisons of contact-based, length-based and duration-based connectivities, not shown in this manuscript, led to little variations for the two case studies considered in this article. However, the use of covariates such as local rainfall and air-mass altitude for defining connectivities, as proposed in Section  \ref{sec:pic}, is expected to potentially impact the inferred networks and deserves to be explored. This perspective would be particularly relevant in the context of aerobiology: e.g., the airborne transport of organic particles, such as bacteria and fungal spores, can be influenced by rainfall favoring the deposition of these particles  \citep{morris2017mapping}.

In statistics, we are not only interested in point estimation, but also in the assessment of estimation uncertainties. In this paper, we however, focused on connectivity estimation, even if quantifying the estimation variance could have been useful for more rigorously investigating temporal variation in connectivities. Formally, the connectivity measures that we defined are integrals. Hence, results on integral numerical approximations (e.g., midpoint, trapezoidal or Monte Carlo integration) can be exploited to assess errors or variances of the connectivity estimates \citep{davis2007methods, caflisch1998monte,geweke1996monte}. However, for this assessment, one should ideally take into account dependencies between connectivity estimates for different pairs of nodes, which is not trivial. Further in-depth methodological developments are required to tackle this issue.

To more finely estimate connectivity, and its uncertainty, one could also take into account, if relevant, the uncertainty about the trajectories themselves. For example, when observed trajectories are smoothed versions of actual trajectories (as it is likely the case for air-mass trajectories calculated with \verb"HYSPLIT") or when the trajectories are partially observed and rather erratic, (i) a probabilistic model grounded on, for instance, a stochastic differential equation, could be used to reconstruct probable trajectories and (ii) the connectivity would be estimated from these reconstructed trajectories. Obviously, step (ii) should incorporate the uncertainty about the trajectory reconstruction impacted by an eventual preliminary step consisting in estimating the parameters of the above-mentioned probabilistic model.

Concerning the application treated in this article, we observed distinct seasonal patterns in the temporal variation of the networks covering the Mediterranean coastline and PACA. In the former case, the networks corresponding to the four clusters shown in Figure \ref{dend_Med}(b-e) exhibit different spatial patterns of hubs (in terms of location and size) and different trends in the main connectivity directions. In the latter case, the differences between the three networks identified with the clustering approach are mostly related to connectivity amplitude. It would be interesting to explore whether this observation made at two very different spatial  scales and resolutions generally holds by studying regions of size similar to PACA all along the Mediterranean coastline.

In the long-term context of our applied research projects connected to aerobiology, the construction and exploration of networks generated by air-mass movements are a way to unravel epidemiological dynamics (and the resulting genetic patterns) of microbial pathogens disseminated at long distance via air movements in the troposphere \citep[see][for a proof of concept]{leyronas2018assessing}. Indeed, even if the pathogen is not explicitly taken into account by the framework proposed in this article, the description of connectivities that it offers provides us a proxy of airborne pathogen movements over long temporal terms and large spatial scales. This proxy is a mean to understand pathogen transportation and to anticipate its long distance dissemination. Specifically, network indices such as those calculated in this article can be associated with particular epidemiological properties such as the probability of long-distance transport of pathogens  \citep{moslonka2011networks,jeger2007modelling,pautasso2014network}. For instance, for plant pathogens, recent studies \citep{nicolaisen2017fungal, bowers2013seasonal, ahospatiotemporal} showed that airborne populations of bacteria and fungi are rather constant across the years, while higher diversity can be observed in different seasons. This statement resonates with our analyses where we observed clear seasonal signals in the estimated monthly spatiotemporal networks in Section \ref{sec:results} whereas the yearly signals were less obvious. 

Finally, the networks estimated using our approach could be a basis for developing epidemiological models (explicitly handling the pathogen) incorporating long-distance dissemination conditional on recurrent air-mass movements. Such models could be exploited to set up surveillance strategies for early warning and epidemic anticipation in order to help reduce the impacts of airborne pathogens on human health, agricultural production and ecosystem functioning \citep{mundt2009aerial}.

\section*{Acknowledgments}

This research was funded by the SPREE project from the French National Research Agency (grant n$^\degree$ ANR-17-CE32-0004-01) and the PHYTOSENTINEL project (grant n$^\degree$ IB-2019-SPE). The authors thank Lo{\"i}c Houde for his technical assistance in the calculation of trajectories with \verb"HYSPLIT".

\bibliographystyle{elsarticle-harv}

\biboptions{authoryear}
\bibliography{main}

\begin{appendices}

\begin{figure}[H]
    \begin{subfigure}[c]{1\linewidth}
        \centering
    \includegraphics[scale=0.55]{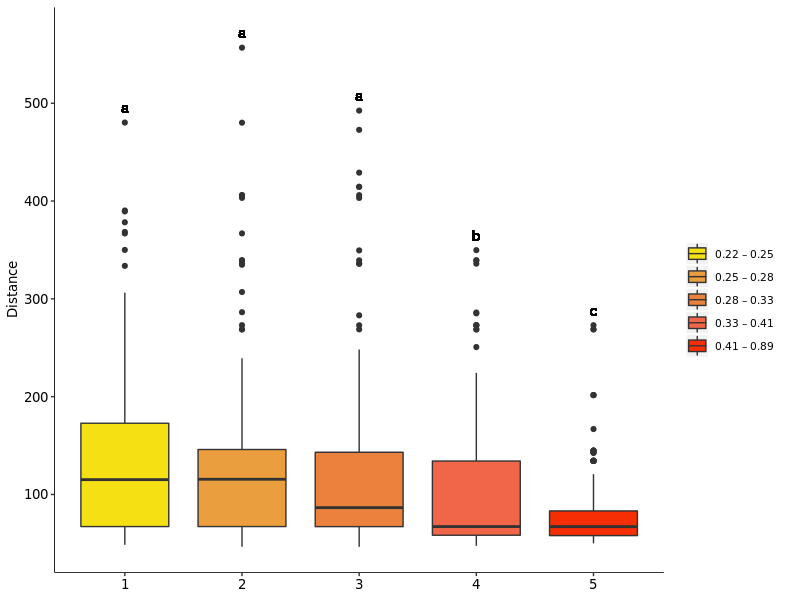}
            \caption{Mediterranean region}
\label{boxplot_Med_dist}
    \end{subfigure}\\[1ex]
    \begin{subfigure}[c]{1\linewidth}
        \centering
    \includegraphics[scale=0.55]{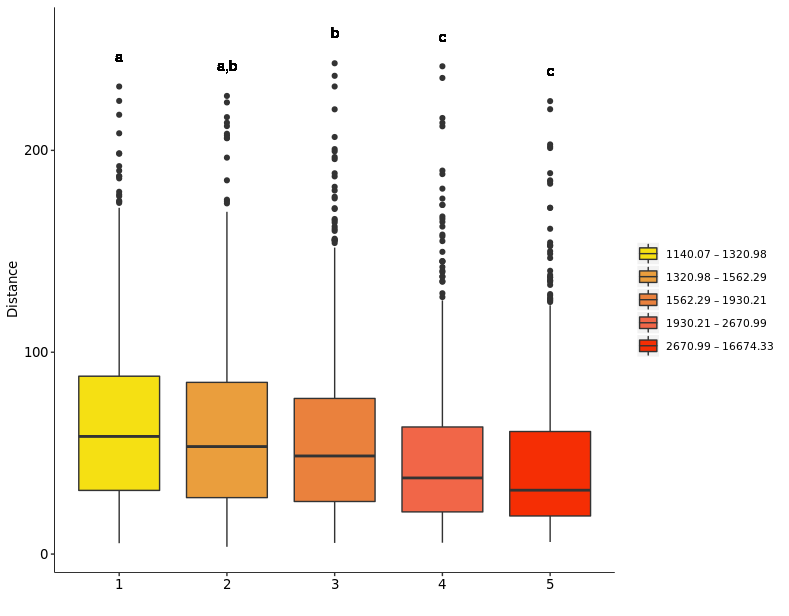}
                \caption{PACA}
        \label{boxplot_PACA_dist}
    \end{subfigure}
    \caption{Boxplot for the distance between the points within the categories. The categories correspond to the intensity of connection between the nodes (a) of the Mediterranean region and (b) PACA. The letters are chosen according to the resulting p-values of Wilcoxon signed-rank test, based on the significance level of 0.05, to compare the distribution of the distances between every couple of the categories in Figure \ref{Boxplot_dist}. The categories having the same letter doesn't have a significant difference between them.}
    
\label{Boxplot_dist}
\end{figure}

\begin{figure}[H]
    \begin{subfigure}[c]{1\linewidth}
        \centering
    \includegraphics[scale=0.6]{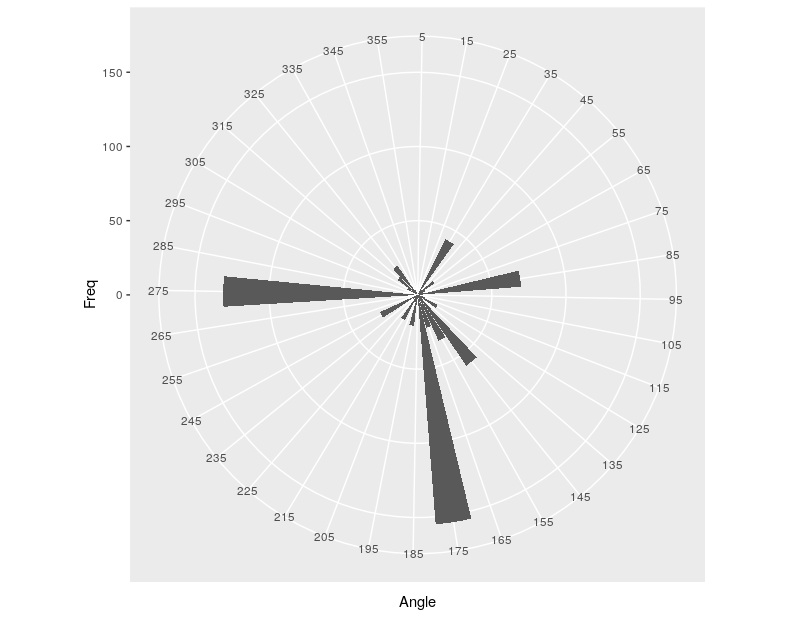}
            \caption{Mediterranean region}
\label{Med_ang}
    \end{subfigure}\\[1ex]
    \begin{subfigure}[c]{1\linewidth}
        \centering
    \includegraphics[scale=0.6]{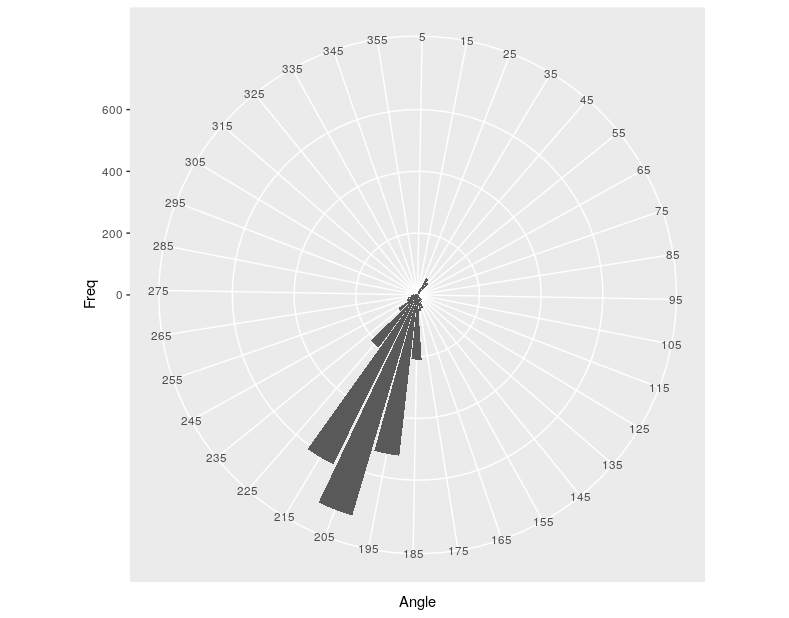}
                \caption{PACA}
        \label{PACA_ang}
    \end{subfigure}
    \caption{Circular histogram illustrating the direction of the connectivities between the nodes whithin the categories (a) of the Mediterranean region and (b) PACA. }
\label{Circular_ang}

\end{figure}

\end{appendices}

\end{document}